\documentclass[sigplan,screen]{acmart}
\usepackage{microtype}
\usepackage{graphicx}
\usepackage{subcaption}
\usepackage{booktabs} % for professional tables

\usepackage{amsmath}
\usepackage{dsfont}
\usepackage{graphicx}
\usepackage{xfrac}
\usepackage{extarrows}
\usepackage{amsbsy}
\usepackage[utf8]{inputenc}
\usepackage{listings,xcolor}
\usepackage[T1]{fontenc}
\usepackage{xcolor}
\usepackage[scaled=0.9]{DejaVuSansMono}
\usepackage{tablefootnote}
\usepackage{array}
\usepackage[toc,page]{appendix}
\usepackage{booktabs}
\usepackage{makecell}
\usepackage{placeins}
\usepackage[compress]{cleveref}
\usepackage{xspace}
\usepackage{mathtools}
\usepackage{algorithm}
\usepackage{algpseudocode}

\DeclareMathOperator*{\argmin}{arg\,min}

\newcommand{\multilinecomment}[1]{}

\usepackage{tikz}% tikz
\DeclareRobustCommand\numcircledtikz[1]{\tikz[baseline=(char.base)]{
    \node[shape=circle,draw,fill,inner sep=1pt] (char)
    {\textcolor{white}{#1}};}}

\usepackage[linewidth=1pt]{mdframed}
\usepackage{lipsum}

\newtoggle{showmarks}
\toggletrue{showmarks}

\iftoggle{showmarks}{
    \newcommand\plan[1]{\textcolor{gray}{#1}}
    \newcommand\todo[1]{\textcolor{blue}{#1}}
    \newcommand\plot[1]{\hfill\newline\textcolor{blue}{[plot] #1\newline}}
    \newcommand\askiad[1]{\textcolor{cyan}{[Athinagoras:] #1}}
    \newcommand\swapnil[1]{\textcolor{blue}{[Swapnil:] #1}}
    \newcommand\myzhao[1]{\textcolor{purple}{[Mark: #1]}}
    \newcommand\christos[1]{\textcolor{magenta}{[Christos:] #1}}
}{
    \newcommand\plan[1]{\unskip}
    \newcommand\todo[1]{\unskip}
    \newcommand\plot[1]{\unskip}
    \newcommand\askiad[1]{\unskip}
    \newcommand\swapnil[1]{\unskip}
    \newcommand\myzhao[1]{\unskip}
    \newcommand\christos[1]{\unskip}
}

\newcommand{\NAME}{ReCycle\xspace}

\AtBeginDocument{%
  \providecommand\BibTeX{{%
    \normalfont B\kern-0.5em{\scshape i\kern-0.25em b}\kern-0.8em\TeX}}}



\setcopyright{acmlicensed}
\acmYear{2024}\copyrightyear{2024}
\acmConference[SOSP '24]{ACM SIGOPS 30th Symposium on Operating Systems Principles}{November 4--6, 2024}{Austin, TX, USA}
\acmBooktitle{ACM SIGOPS 30th Symposium on Operating Systems Principles (SOSP '24), November 4--6, 2024, Austin, TX, USA}
\acmDOI{10.1145/3694715.3695960}
\acmISBN{979-8-4007-1251-7/24/11}

\begin{document}

\title{\NAME: Resilient Training of Large DNNs using Pipeline Adaptation}

\author{Swapnil Gandhi}
\email{gandhis@stanford.edu}
\affiliation{\textit{Stanford University}\country{}}

\author{Mark Zhao}
\email{myzhao@cs.stanford.edu}
\affiliation{\textit{Stanford University}\country{}}

\author{Athinagoras Skiadopoulos}
\email{askiad@stanford.edu}
\affiliation{\textit{Stanford University}\country{}}

\author{Christos Kozyrakis}
\email{kozyraki@stanford.edu}
\affiliation{\textit{Stanford University}\country{}}

% \renewcommand{\shortauthors}{Gandhi, et al.}

\begin{abstract}
Training large Deep Neural Network (DNN) models requires thousands of GPUs over the course of several days or weeks. At this scale, failures are frequent and can have a big impact on training throughput. Utilizing spare GPU servers to mitigate performance loss becomes increasingly costly as model sizes grow. \NAME is a system designed for efficient DNN training in the presence of failures, without relying on spare servers. It exploits the inherent {\it functional redundancy} in distributed training systems -- where servers across data-parallel groups store the same model parameters -- and {\it pipeline schedule bubbles} within each data-parallel group. When servers fails, \NAME dynamically re-routes micro-batches to data-parallel peers, allowing for uninterrupted training despite multiple failures. However, this re-routing can create imbalances across pipeline stages, leading to reduced training throughput. To address this, \NAME introduces two key optimizations that ensure re-routed micro-batches are processed within the original pipeline schedule's bubbles. First, it decouples the backward pass into two phases: one for computing gradients for the input and another for calculating gradients for the parameters. Second, it avoids synchronization across pipeline stages by staggering the optimizer step. Together, these optimizations enable adaptive pipeline schedules that minimize or even eliminate training throughput degradation during failures. We describe a prototype for \NAME and show that it achieves high training throughput under multiple failures, outperforming recent proposals for fault-tolerant training such as Oobleck and Bamboo by up to $1.46\times$ and $1.64\times$, respectively. 
\end{abstract}

% \keywords{Do, Not, Us, This, Code, Put, the, Correct, Terms, for,


\maketitle

\section{Introduction}\label{sec:intro}
Deep Neural Networks (DNNs) are consistently achieving milestone results in domains such as natural language processing~\cite{chatgpt}, speech recognition~\cite{speech-recognition}, and computer vision \cite{imagenet}.
%with performances that now closely match human levels of expertise~\cite{gpt-4}. 
Since the rapid growth of DNNs is a major contributor to recent breakthroughs~\cite{model-size-growing}, we are now in a global race to develop increasingly-large foundation models~\cite{foundation-models, chinchilla}.
Both proprietary~\cite{gpt-3, gpt-4, gemini} and open-source~\cite{llama-2, dbrx} models have tens to hundreds of billions of parameters. Training such models requires uninterrupted access to thousands of accelerators (e.g., GPUs) for several weeks~\cite{megatron-nlg, megascale, training-at-meta}. DNN training is essentially a large supercomputing job that relies on hybrid parallelism, which concurrently leverages data, tensor, and pipeline parallelism strategies~\cite{alpa}.
%Data , tensor , and pipeline parallelism techniques are used concurrently~\cite{alpa}.  %\askiad{I can update this paragraph}

%This progress, particularly noticeable in areas like Natural Language Processing (NLP)~\cite{llama, llama-2, opt, gpt-2, gpt-3, chatgpt, gemini, gshard, switch-transformer}, owes much to the upscaling of model sizes, the expansion of training datasets, and enormous computational budget~\cite{model-size-growing, pathways, chinchilla, megatron-nlg, training-at-meta, scaling-laws-for-nlp}. For example, in a span of just six years, the computational power required for model training has surged by a factor of 300,000~\cite{rising-compute-power-needs, ai-memory-wall}. The task of training these colossal models, e.g. Switch Transformer~\cite{switch-transformer} has 1.6 Trillion parameters, far exceeds the memory limits of a single accelerator and presents a substantial challenge, typically necessitating months of uninterrupted training over a vast array of high performance accelerators, thereby garnering significant attention within the field~\cite{tensorflow, pytorch-fsdp, gpipe, chimera, hanayo, zero, zero-offload, opt, megatron-lm, sequence-parallelism, activation-checkpointing}. To address these rigorous demands, distributed hybrid-parallel training~\cite{megatron-sc, alpa, amazon-sagemaker, pytorch-distributed, pathways} has become the standard approach, effectively blending model and data parallelism to efficiently tackle the complexities involved in scaling up training processes.

As the scale and duration of training increases, so does the likelihood of encountering failures~\cite{failures-in-large-scale-systems,microsoft-training-trace,megascale,check-n-run,mlaas-in-the-wild,tpu-resiliency}. While scale-out workloads, such as distributed databases and analytics, are designed to work around server failures, this is not yet the case for DNN training. The rigid parallelization of DNN training implies that each failure can set off a domino effect.
All resources are forced to idle while a failed node is replaced and the job is re-optimized and re-started on the new hardware configuration. For example, the training of OPT-175B included 178,000 GPU-hours of wasted time due to various malfunctions~\cite{bloom}. 

%Each failure sets off a domino effect, causing all participating computational resources to idle as they await the repair or substitution of the compromised resource. This inefficiency leads to a dramatic under-utilization, as evidenced by the loss of roughly 178,000 GPU hours in the training of the OPT-175B model due to various malfunctions~\cite{bloom}. It is crucial to recognize that such downtime not only wastes valuable compute time but also imposes a substantial economic cost, largely due to the scarcity and high expense associated with AI accelerators~\cite{bloom-hf-blog,laion}.

Fault tolerance in DNN training involves three issues: {\it fault detection}, {\it checkpointing}, and {\it efficient execution in the presence of faults}. Large-scale training systems implement comprehensive monitoring of the health of hardware and software components, so that faults or stragglers are quickly detected and diagnosed~\cite{megascale}. Recent research has focused on reducing the overhead of periodic checkpoints of the training job state~\cite{check-n-run, checkfreq, activation-checkpointing}. Efficient execution in the presence of faults has received less attention. A common approach in industry is to maintain a reserve of spare GPU servers that will replace failed servers when needed. While conceptually simple, this approach is expensive at scale. As the frequency of faults increases, so does the number of spares. 

The alternative approach is to {\it continue training with the subset of resources available}. This is similar to how scale-out systems approach resilience to failures~\cite{mapreduce}. Data parallelism provides one simple implementation as it creates full replicas of the model across groups of GPUs. When one GPU server fails, we take all the servers in its data-parallel group offline and continue training the remaining model replicas. Unfortunately, this approach increases the blast radius of the failure and its impact to performance. When training a 530B GPT model using the hybrid-parallel scheme of Megatron-LM~\cite{megatron-sc}, a single node failure would force 280 GPUs to go offline and cause a roughly 11\% drop in training throughput. 

Two recent projects have exploited {\it pipeline parallelism} instead. 
Bamboo~\cite{bamboo} relies on redundant computation and executes each pipeline stage on two nodes even in the fault-free case. When a node fails, the alternative nodes for its stages will do the work. Bamboo's drawback is the drop in fault-free training throughput due to the redundant computations and added memory pressure on each GPU node. Oobleck~\cite{oobleck} avoids overheads in fault-free execution by precomputing a number of templates for pipeline parallelism that use a different number of nodes. As nodes fail, Oobleck replaces the original pipeline template with one that uses fewer nodes. The performance challenge for Oobleck is balancing work across heterogeneous pipelines as the pipelines with fewer nodes can become stragglers. Both Bamboo and Oobleck require significant re-configuration as nodes fail or rejoin, which can be a challenge as failure rates increase.

In this paper, we introduce {\it \NAME}, a novel scheme for resilient training with hybrid-parallel systems. \NAME enables efficient execution in the presence of failures, without the need for spares and without any impact on model accuracy. Similar to Bamboo and Oobleck, we exploit pipeline parallelism. But unlike these previous proposals that handle failures within a single pipeline,  \NAME utilizes the inherent \emph{functional redundancy across pipelines} in hybrid-parallel training systems. \NAME re-routes the micro-batches of failed nodes to \emph{peer nodes} that process the same pipeline stage for the model in the other data-parallel groups. Peer nodes store the same model parameters and can handle the extra work without the need for parameter re-shuffling.  

If done naively, \NAME's {\it adaptive pipelining} increases training time and memory usage due to the additional work for the peers of failed nodes.
\NAME uses a series of optimizations that exploit unique characteristics of hybrid-parallel training to eliminate all or most of the inefficiencies. \NAME uses the \emph{bubbles} in the pipeline schedule of peers to execute the additional work at low or no overhead. These bubbles typically appear during the \emph{start-up} and \emph{cool-down} phase of the pipeline schedule, but not during the \emph{steady-state} when most additional work must occur. \NAME overcomes this challenge by {\it decoupling back propagation} for micro-batches into two distinct gradient computations, one relative to the input and one relative to the parameters. %, one focusing on gradient computation relative to the input and one and one updating model parameters.
The two steps are independently scheduled in order to better leverage bubbles in the the \emph{cool-down} portion of the pipeline schedule.
The selective application of decoupled back propagation exploits the imbalance in memory usage across stages in hybrid-parallel training while avoiding an increase in peak memory pressure. Finally, \NAME staggers the execution of the optimizer tasks across pipeline stages in order to leverage bubbles from the \emph{start-up} phase of the pipeline schedule for the next training iteration. % to the \emph{cool-down} phase of the pipeline schedule for the current iteration.

%To minimize the time overhead associated with \emph{surrogates}, \NAME capitalizes on the presence of \emph{bubbles}—inherent computation stalls occurring due to inter-stage dependencies within systems that employ synchronous 3D parallelism. 
%These bubbles typically appear during the \emph{start-up} and \emph{cool-down} phases of an iteration, but not during the \emph{steady-state}. \NAME strategically splits the backward computation into two distinct parts: one focusing on computing the gradient for the input and the other for the model parameters, thus utilizing these bubbles to smooth out computational peaks.

%Moreover, the process of handling \emph{surrogates} inevitably increases memory demands. \NAME addresses this by leveraging another key insight: the existence of memory imbalances across the stages of a pipeline in 3D parallelism. By utilizing this surplus memory capacity, \NAME efficiently accommodates the additional memory requirements needed to process \emph{surrogates}, ensuring that the system remains balanced and effective despite the shifts in workload caused by node failures.

We implemented \NAME on top of the DeepSpeed training framework~\cite{deepspeed}. % using components from Megatron-LM~\cite{megatron-lm}.
The key component of \NAME is the {\it Planner}, which utilizes dynamic programming and mixed-integer linear programming (MILP) to determine adaptive pipeline schedule in presence of multiple failures. Operating offline, the {\it Planner} precomputes efficient schedules for a predefined number of failures, which are then applied as necessary during training.
We evaluated \NAME using DNNs with billions of parameters like GPT-3~\cite{gpt-3}, using both real-world experiments and simulations of large training systems. We show that \NAME supports high throughput training even at high failure counts, e.g. $\geq$10\% of the overall system. \NAME outperforms Oobleck by $1.46\times$ in training scenarios with realistic GPU failures.
\NAME's advantage stems from its efficient scheduling of re-routed work and its ability to avoid major re-shuffling of model parameters upon failures. \NAME outperforms Bamboo by $1.64\times$. It is also able to efficiently train much larger models than Bamboo, which requires large memory overheads for its fault tolerance mechanism. %  efficiently  

\section{Background and Motivation}\label{sec:background}

%\christos{commented out this intro}
%In this section, we briefly introduce hybrid parallelism that is commonly used for large model training. We also discuss existing fault tolerance strategies for hybrid-parallel training and highlight their limitations.

\subsection{Distributed DNN Training}

State-of-the-art {\it Deep Neural Network (DNN)} models consist of tens to hundreds of billions of parameters and are trained on datasets with many trillions of tokens~\cite{gpt-3, gpt-4, llama-2, bloom, opt, switch-transformer, gshard}. Their training requires thousands of GPUs for days or weeks~\cite{chinchilla, megatron-nlg, training-at-meta, scaling-laws-for-nlp}. For example, Llama-3 was trained on 15 trillion tokens, using two clusters of 24K GPUs~\cite{llama-3}. 

DNN training is parallelized using three primary forms of parallelism. {\it Data parallelism} (DP) processes subsets of the input data in parallel across GPU groups, each of which stores the entire model~\cite{horovod, imagenet-in-hours, imagenet-in-minutes}. 
%In addition to significant memory demands, 
Data parallelism requires high network bandwidth for the all-reduce operations that reconcile model parameters across all replicas at the end of iteration steps. 
Tensor and pipeline parallelism are forms of {\it model parallelism} that facilitate training large models by sharding the model across GPUs. {\it Tensor parallelism} (TP) partitions the parameters of each layer across GPUs in order to parallelize each linear algebra operation within a layer~\cite{megatron-lm}. It incurs high communication costs that are not easily hidden by computation due to frequent
%\christos{I am not sure if it is all-reduce here. it is all-to-all and reduce-scatter right?}\swapnil{Megatron-LM has special handcrafted partitioning such that it runs two all-reduces, one for operator \(f\) in forward pass and one for operator \(g\) in backward pass.} 
all-reduce operations in both the forward and backward pass. 
{\it Pipeline parallelism} (PP) divides the model into sequential groups of layers or {\it stages}. Micro-batches of data are processed in parallel in a pipelined manner across these stages~\cite{gpipe, pipedream, chimera, hanayo, bpipe}. Pipeline parallelism requires lower network bandwidth as stages only communicate activations and gradients at layer boundaries. However, it achieves lower compute utilization as the number of stages increases due to pipeline dependencies and bubbles (idle slots) in the pipeline schedule.  

\begin{figure}
    \centering
    \includegraphics[width=1\linewidth]{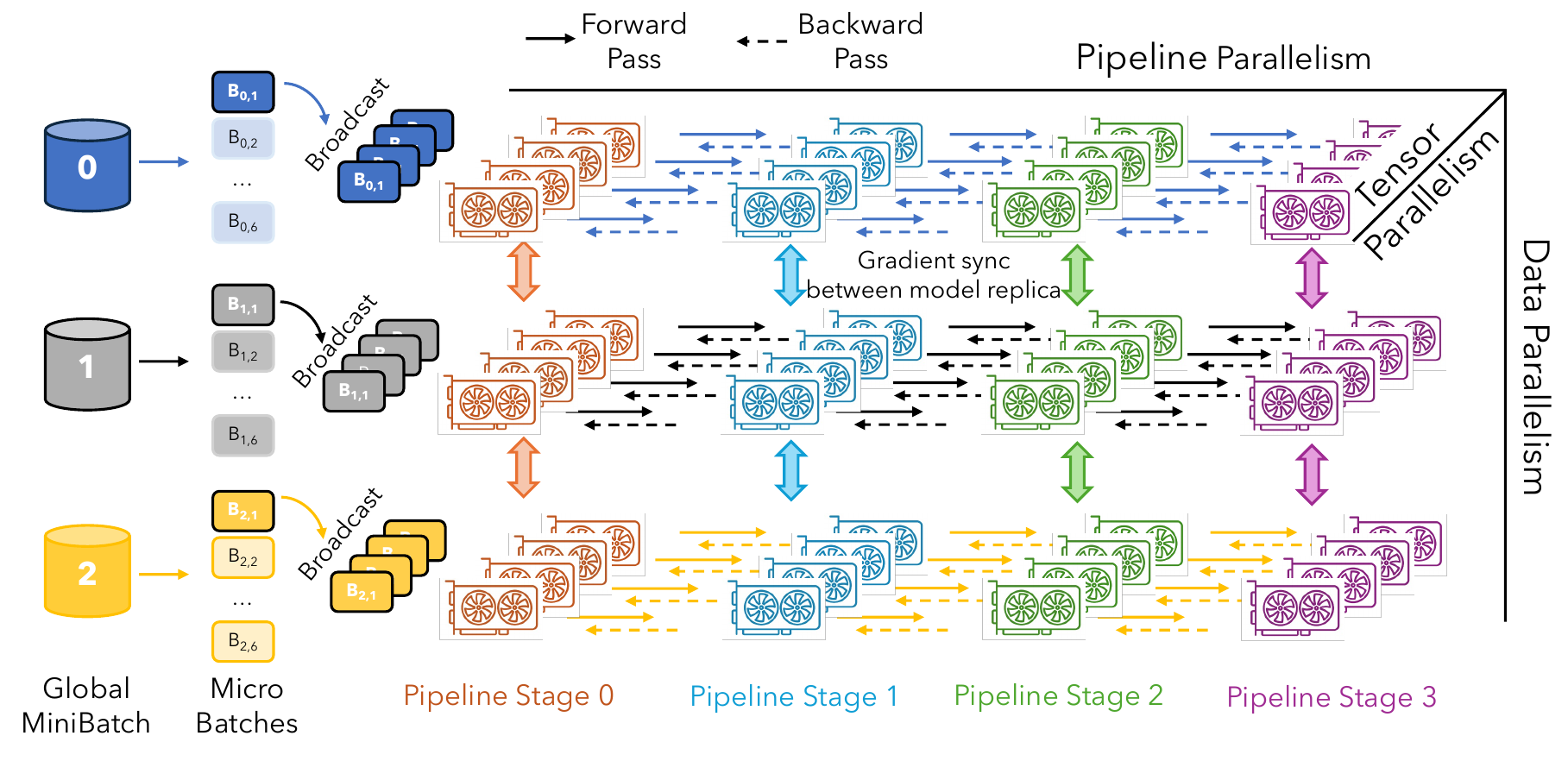}
    \caption{Illustration of hybrid parallelism.  Pipeline stages are denoted with different colors. Within each pipeline stage, operators are partitioned through tensor parallelism. The global batch is split into micro-batches across pipelines.} % The input global mini-batch is split into discrete micro-batches.} % \(B_{i,j}\) is the \(j^{th}\) micro-batch for the \(i^{th}\) data parallel replica.}
    \label{fig:hybrid-parallelism}
\end{figure}

Large-scale training systems balance these trade-offs by utilizing all three forms of parallelism as shown in Figure~\ref{fig:hybrid-parallelism}~\cite{megatron-sc, chimera, hanayo, activation-checkpointing, deepspeed, pytorch-distributed, pytorch-fsdp, alpa}. This is known as {\it hybrid-parallelism}. In the most common case, systems such as DeepSpeed~\cite{deepspeed} or Alpa~\cite{alpa} use a combination of tensor parallelism within a multi-GPU server, pipeline parallelism across multi-GPU servers, and data parallelism across pipelines. Hybrid parallelism enables large model training with graceful scaling and reasonable training times (weeks to few months) using  optimized clusters with 1000s of densely-connected GPUs.

%For example, \christos{take one paper like Megatron or MEgascale and explain how the degree of parallelism in each dimension}

%As DNNs increase in size and are trained on ever-larger datasets, relying solely on data parallelism or model parallelism often proves insufficient for efficient training. Data parallelism involves distributing input data across multiple GPUs, each of which must store the entire model, imposing significant memory demands. Model parallelism, on the other hand, facilitates the training of large models by distributing different parts of the model across multiple GPUs. For instance, pipeline parallelism divides the model into sequential groups of layers, or stages, while tensor parallelism splits each layer into smaller tensor segments. However, pipeline parallelism introduces inefficiencies such as pipeline bubbles that reduce compute utilization as the number of stages increases. Tensor parallelism, meanwhile, incurs high communication costs that are not easily offset by computation due to frequent all-reduce operations essential for both the forward pass and back-propagation. To address these challenges, a blend of data and model parallelism--often referred to as hybrid parallelism--is typically employed to effectively train large-scale DNN models.

\subsection{Fault Tolerance in Distributed DNN Training} 

Large training systems include thousands of GPUs, CPUs, memory chips, networking chips, cables of various types, and power conversion and cooling devices. Scale allows for high performance but also leads to frequent faults ranging from software errors to full hardware malfunctions. The Mean Time Between Failure (MTBF) for large training systems can be minutes~\cite{failures-in-large-scale-systems, mlaas-in-the-wild, megascale, unicorn}. For example, large-scale training clusters at Microsoft see a failure every $\approx45$ minutes~\cite{microsoft-training-trace}.

%a recent study of large-scale DNN training clusters at Microsoft reports the mean time between job failures as $\approx45$ minutes~\cite{microsoft-training-trace}. %, excluding early failures~\cite{microsoft-training-trace}.

In contrast to scale-out frameworks like Map-Reduce~\cite{mapreduce}, which quickly adjust to dynamic changes in resource availability, DNN training operates like a supercomputing job. It relies on fixed sharding and parallelization strategies and gang-scheduled execution. All computational resources must concurrently run uninterrupted, making the system highly susceptible to disruptions from any failure. For example, Meta encountered over 100 hardware failures while training OPT-175B, resulting in the loss of 178,000 GPU-hours~\cite{bloom,bloom-hf-blog}. Similar failure rates have been reported by ByteDance~\cite{megascale}, Alibaba~\cite{unicorn}, LAION~\cite{laion}, Microsoft~\cite{microsoft-training-trace} and Google~\cite{tpu-resiliency}. At this scale, faults are regular occurrences that require systematic optimizations to ensure resilient and efficient training~\cite{unicorn, megascale, microsoft-training-trace, mlaas-in-the-wild}.

%As Deep Neural Network (DNN) models grow in complexity and size, the infrastructure needed to train these models requires an unprecedented scale of hardware resources, including conventional compute units, accelerators, and extensive networking and storage systems. Unfortunately, the expansion in system components and their complexity also leads to an increase in the frequency of faults, which range from hardware malfunctions to software errors, drastically reducing the Mean Time Between Failures (MTBF) to mere hours or even minutes. In contrast to flexible distributed applications like Map-Reduce, which can adjust to dynamically changing resource allocations, DNN training usually depends on fixed sharding strategies and gang-scheduled executions. This necessitates that all computational resources function concurrently and uninterrupted, thus making the system highly susceptible to disruptions from any single point of failure. For example, during the training of OPT-175B, Meta's AI team encountered over 100 hardware failures, necessitating more than 100 significant restarts. Therefore, it is imperative to recognize that such faults are not anomalies but regular occurrences that require systematic management to ensure the resilience and efficiency of training operations.

Enhancing fault tolerance in distributed training requires addressing three critical areas: fast error detection, checkpointing, and efficient execution in presence of faults.

\subsubsection{Error Detection} Fast Error Detection is vital for minimizing downtime and preventing the propagation of errors. Effective error detection mechanisms include both hardware and software solutions. Hardware mechanisms, such as error correction codes (ECC), provide immediate detection and correction of data corruption. Software solutions can involve timeouts and heartbeat signals to identify system failures quickly. Additionally, some errors, particularly those that are silent or not immediately apparent, are detected through irregularities in the loss function. These anomalies, often indicative of silent data corruptions, can significantly impact training outcomes if not addressed promptly~\cite{laion, sdc-google, sdc-meta}.

\subsubsection{Checkpointing} Checkpointing plays a crucial role in fault tolerance by allowing the system to recover from a recent stable state rather than restarting the entire training process. It involves saving the model's state at regular intervals. However, naive checkpointing methods that write data to remote storage can introduce significant pauses in the training process due to latency and bandwidth constraints. To address this, extensive research has focused on optimizing checkpointing techniques to reduce overhead and improve efficiency~\cite{checkfreq, check-n-run, gemini, megascale}, as detailed in Section~\ref{sec:related}.

\subsubsection{Efficient DNN Training in the Presence of Failures}\label{sec:background_systems}
Our work focuses on ensuring that the system remains operational even when parts of it fail. 
%Our work focuses on it. %Once a failure is detected and isolated and a proper checkpoint identified to resume from, we need to continue training at the best possible performance and/or lowest possible cost.  
A common method is to use \textit{warm} spares, preloaded with model parameters to reduce recovery time. While this ensures that training efficiency returns to pre-failure levels, it becomes costly as systems scale and fault frequency rises. Spare servers need similar network bandwidth as the ones they replace, adding networking costs and requiring coarse-grained allocation~\cite{google-ocs}. For example, training a 530B GPT model requires 280 spare GPUs, provisioned according to the full capacity of a data parallel group, increasing costs by 11\%~\cite{megatron-sc}. Use of \textit{cold} spares to run small jobs can offset some of these costs, however eviction delays upon failure can cause significant stalls for large jobs.

%Ensuring continual training in presence of one or more failures is crucial for maintaining the resilience and efficiency of training large-scale Deep Neural Networks (DNNs). We evaluate several strategies for managing failures, each with its unique set of advantages and drawbacks.

%Lastly, robust execution under faults entails ensuring the system remains operational even when parts of it fail. This is achieved through strategies like task redundancy and dynamic task reallocation, which help sustain operational throughput. Addressing these aspects is vital for developing a durable distributed training framework capable of managing inevitable system failures effectively.

\begin{figure}
    \centering
    \includegraphics[width=1\linewidth]{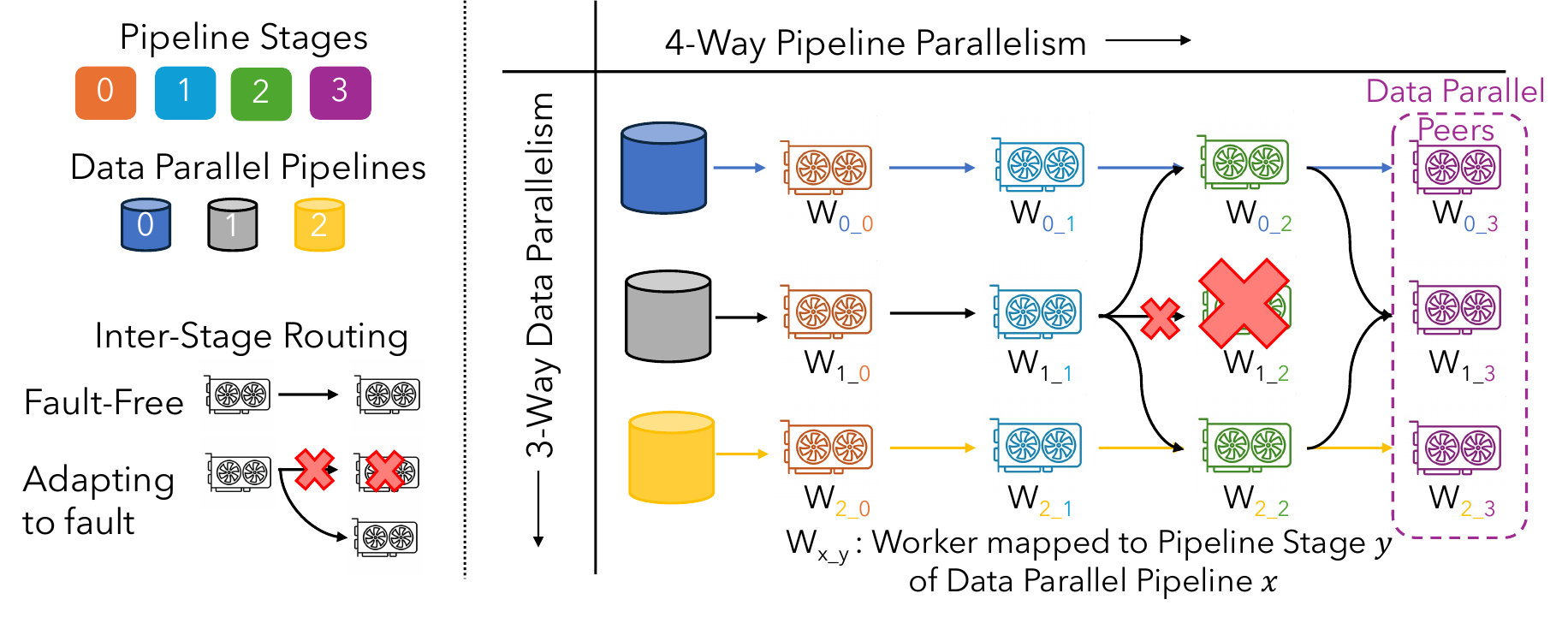}
    \caption{Adaptive pipelining when \(W_{1\textunderscore2}\) fails.  The micro-batches from worker \(W_{1\textunderscore1}\), originally intended for \(W_{1\textunderscore2}\), are dynamically re-routed to workers \(W_{0\textunderscore2}\) and \(W_{2\textunderscore2}\), ensuring that the training process continues without interruption.}
    \label{fig:adaptive-pipeline}
\end{figure}
\begin{figure*}[t!]
    \centering
    \begin{subfigure}[c]{\textwidth}
        \centering
        \includegraphics[height=1.4cm]{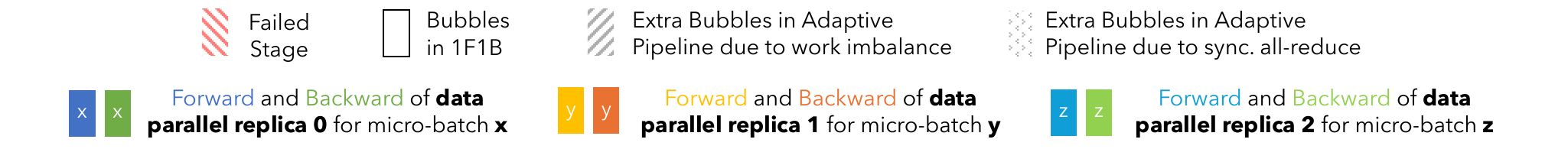}
    \end{subfigure}
    \hfill
    \begin{subfigure}[c]{0.4\linewidth}
        \centering
        \includegraphics[width=\linewidth]{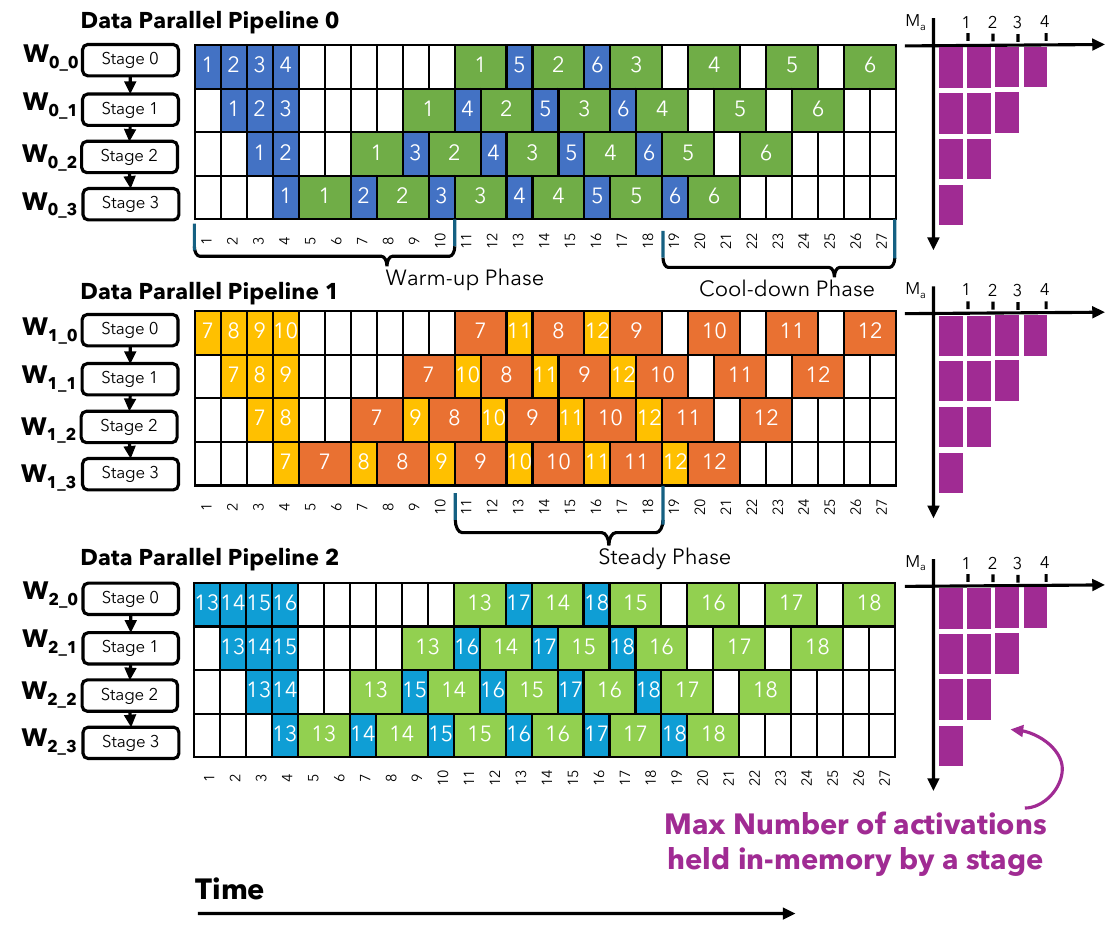}
        \caption{Fault-Free 1F1B Schedule}
        \label{fig:fault-free-schedule}
    \end{subfigure}
    \hfill
    \begin{subfigure}[c]{0.52\linewidth}
        \centering
        \includegraphics[width=\linewidth]{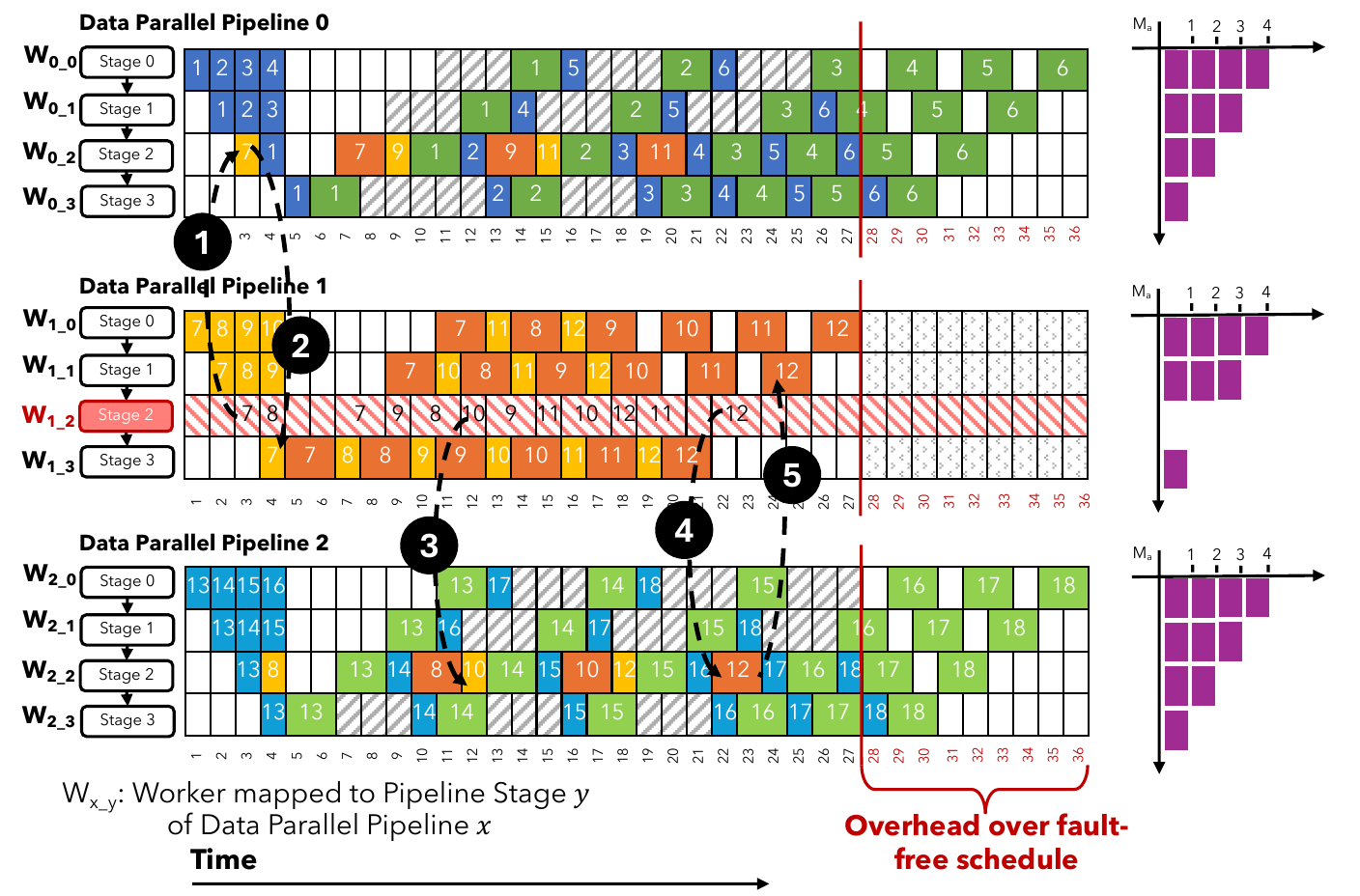}
        \caption{Adaptive Schedule when \(W_{1\textunderscore2}\) fails.}
        \label{fig:adaptive-schedule}
    \end{subfigure}
    \caption{Hybrid-parallel training across 12 workers with 3 data-parallel pipelines, each with 4 pipeline stages, and 6 micro batches. ~\ref{fig:fault-free-schedule} shows a 1F1B fault-free schedule. ~\ref{fig:adaptive-schedule} shows how \emph{Adaptive Pipelining} 
   re-routes micro-batches from failed worker \(W_{1\textunderscore2}\) to its functional peers \(W_{0\textunderscore2}\) and \(W_{2\textunderscore2}\).}
    \label{fig:fault-free-and-adaptive-schedule}
\end{figure*}

An alternative approach is to continue training with the largest subset of available resources. Ideally, if \(x\%\) of GPU servers fail, training proceeds at no less than \((100 - x)\%\) of fault-free throughput. A simple method, known as elastic batching, drops an entire data parallel group when a node fails, allowing the remaining groups to continue with the same parallelization~\cite{dynamic-data-dropout, dynamic-mini-batching}. However, this approach has significant drawbacks, as a single node failure can take \(PP \times TP\) nodes offline, reducing throughput by \(1/DP\) (e.g., 17\% for a 1T parameter GPT model~\cite{megatron-sc}). Moreover, this approach requires reducing the batch size or recompiling the program to optimize each data parallel group for a larger batch portion, disrupting the balance and efficiency of the training job.

%280 GPUs in the 530B GPT example mentioned above).
%. For instance, in the case of the 530B GPT model mentioned earlier, the failure of one node leads to 280 GPUs being taken offline, which in turn reduces the training throughput by 11\%~\cite{megatron-sc}.
%The throughput drop spikes to 17\% for the  1000B GPT model. 
%Moreover, this requires reducing the training batch size or recompiling the program to optimize each data parallel group for a larger portion of the batch. This can disrupt the balance and efficiency of the entire training job.

Two recent projects use pipeline parallelism for fault-tolerant training without spares. Bamboo~\cite{bamboo}, drawing inspiration from RAID~\cite{raid}, employs redundant computation (RC) where each pipeline stage is replicated on two nodes, even in fault-free cases. When a node fails, its backup handles its and forward and backward passes of failed node. Though RC hides some overhead in pipeline bubbles, it still significantly reduces throughput (see Section~\ref{sec:evaluation}) and increases GPU memory pressure, limiting scalability. Moreover, Bamboo must restart with a full reconfiguration from a checkpoint for as few as two adjacent node failures. 

Oobleck~\cite{oobleck} uses pipeline parallelism to ensure resilient execution with no overhead when no faults occur. It utilizes precomputed templates for pipeline parallelism, each differing in stages, micro-batch configurations, and node counts. If a pipeline encounters a failed node, it switches to a template with fewer nodes, which may reduce training throughput. To prevent a slow pipeline from affecting the overall job, Oobleck distributes the global mini-batch based on compute power, leading to increased training time. Pipeline re-configuration when nodes fail or re-join can also become an overhead if failures are frequent. Additionally, Oobleck treats each pipeline as a \emph{black-box} and does not leverage bubbles to mitigate overheads from node failures. 

\section{\NAME Techniques}\label{sec:system}

\NAME aims to support efficient distributed training in the presence of faults. It requires no spare servers and has no impact on model accuracy compared to fault-free training. \NAME can tolerate multiple hardware failures and maintains training throughput proportional to the number of functional servers available. \NAME is optimized for fast recovery as failures do not require significant re-shuffling of model parameters between functional nodes. 

This section reviews the key \NAME techniques: \textit{Adaptive Pipelining}, \textit{Decoupled BackProp}, and a \textit{Staggered Optimizer}. Section~\ref{sec:design} presents the \NAME system design. 

%In contrast to current solutions, which are forced to sacrifice training progress, model accuracy, or cost efficiency \myzhao{revise to reflect take-away in 2 if this is not accurate}, \NAME proposes a new approach to enabling resilient distributed DNN training.
%Specifically, \NAME is designed to continue training, even in the face of multiple faults, with minimal throughput degradation compared to the fault-free case.
%\NAME does not require checkpoint resumptions in the vast majority of failure modes, does not impact model accuracy, and does not depend on idle hot standbys.

\subsection{Adaptive Pipelining: Working Around Failures}
\label{sec:adaptive-pipeline}

\NAME exploits two key properties in hybrid-parallel training systems. First, there is \textit{functional redundancy} across data-parallel pipelines.
%pipeline stages are \emph{functionally redundant}. %\askiad{A first read implies redundancy across PP and not the DP degree. Something like: "Data parallelism adds functional redundancy to each pipeline stage".}
%Across data parallel pipelines, t
The GPUs that process the same pipeline stage hold identical parameters and only differ in the micro-batches they process. In Figure~\ref{fig:adaptive-pipeline}, for example, workers \(W_{0\_2}\),  \(W_{1\_2}\), and \(W_{2\_2}\) are {\it peers} that hold identical parameters for stage 2 of the 4-stage pipeline. Second, there are \textit{bubbles (idle slots) in the pipeline schedule} for each worker. The 1F1B pipeline schedule~\cite{pipedream} in Figure~\ref{fig:fault-free-schedule} has 9 bubbles in the repeating 27-slot schedule for worker \(W_{0\_2}\).

%\christos{We have not really defined data parallel group. It can be confusing with data parallel pipeline in the text. So I am switching to peer group which seems more intuitive based on the definition above. Correct it if needed. Needs to propagate to intro}

\textit{Adaptive Pipelining} exploits functional redundancy and bubbles by \textit{dynamically re-routing micro-batches} from a failed worker to its functioning peers in other data-parallel pipelines. We aim to use the bubbles in peers' schedules to process the micro-batches for the failed worker with a low performance penalty.  We evenly distribute micro-batches across all functional peers. 
Since all pipelines perform a synchronized all-reduce at the end of each iteration, load balancing ensures that no single worker (and thus pipeline) is overloaded, delaying the progress of the entire training iteration. 
%\askiad{Clarify why all-reduce helps.}

Figure~\ref{fig:adaptive-pipeline} shows an example with three data parallel, 4-stage pipelines. Worker \(W_{1\_2}\) fails. %\askiad{I'd move this sentence in the first paragraph of 3.1.}
Upon detecting the failure, \textit{Adaptive Pipelining} will redirect its input to workers \(W_{0\_2}\) and \(W_{2\_2}\). These two peer workers will process micro-batches received from worker \(W_{1\_1}\) in addition to their regular pipeline load.
The output for these additional micro-batches will be sent back to worker \(W_{1\_3}\), the original recipient of the output of the failed worker \(W_{1\_2}\). All other workers operate as in the fault-free schedule. In essence, \textit{Adaptive Pipelining} repairs the functionality of pipeline 1 by exploiting bubbles in the data parallel peers of failed worker \(W_{1\_2}\). Note that {\it Adaptive Pipelining} requires {\it no model parameter re-shuffling or re-partitioning across workers} in order to resume after a failure.
%All workers across all pipelines retain the state of the pipeline stage they originally served.
Hence recovery is fast, unlike schemes like Oobleck that require re-configuring an entire pipeline upon a failure~\cite{oobleck}. % \askiad{Unclear during a first-read which pipeline is shorter and why}. 

Figure~\ref{fig:fault-free-and-adaptive-schedule} shows the detailed 1F1B schedule for the system in  Figure~\ref{fig:adaptive-pipeline}. After the failure, the \emph{forward pass} micro-batch 7 for failed worker \(W_{1\_2}\) is re-assigned to its functional peer  \(W_{0\_2}\) at time step 3 (\numcircledtikz{1}). The subsequent output is forwarded back to worker \(W_{1\_3}\) at time step 4 (\numcircledtikz{2}).
Similarly, the \emph{forward pass} of micro-batch 10 is re-routed to worker \(W_{2\_2}\) at time step 12 (\numcircledtikz{3}).
The \emph{backward pass} operates similarly in the opposite direction. 
Worker \(W_{2\_2}\) receives  from worker \(W_{1\_3}\) the gradients for micro-batch 12 at time step 22 (\numcircledtikz{4}). The output gradients are passed to worker \(W_{1\_1}\) at time step 24 (\numcircledtikz{5}).
%\askiad{I would remove step 3. It takes the reader time to understand what's going on and why batch 10 and not 9. I would explain fwd for 7 (step 1, 2), bwd for 12 or another one, and then a sentence that says; in the end, forward and backward for all micro-batches 7,8,9,10,11,12 are treated similarly.}
The overall mathematical computation remains {\it unchanged} from the fault-free 1F1B schedule, ensuring that \textit{Adaptive Pipelining} does not impact model convergence.

{\bf Are sufficient bubbles available?} 
\emph{Adaptive Pipelining} exploits the \(3 \times (PP-1) \times DP\) existing but previously unused bubbles in each pipeline to recover from failures.
For example, the 405 billion parameter LLaMA-3 is trained with synchronous 1F1B over \(8192\) GPUs using hybrid parallelism with tensor, pipeline, and data parallelism degree~\(TP=8\), \(PP=16\), and \(DP=64\), respectively~\cite{llama-3}.
\NAME can leverage \((3 \times (16-1) \times 64) = 2880\) bubbles per iteration to accommodate \( \frac{2880}{3} = 960\) rerouted micro-batches per iteration. For a training job with global batch size of 2048, this is sufficient to handle 30 simultaneous failures.

{\bf The scheduling challenge:} 
%Adaptive Pipelines can introduce bubbles.
%\christos{feel free to improve} 
While sufficient bubbles exist, these bubbles are concentrated in the warm-up and cool-down phase of the 1F1B schedule. 
For example, there are 18 idle slots across data-parallel peers of failed node \(W_{1\_2}\) in Figure~\ref{fig:fault-free-schedule} to accommodate its micro-batches.
\textit{Adaptive Pipelining} needs to reroute micro-batches to peer workers \(W_{0\_2}\) and \(W_{2\_2}\) mostly in the middle of their steady-state schedule, which is optimized to be bubble-free for efficiency~\cite{pipedream}.
As shown in Figure~\ref{fig:adaptive-schedule}, this has a significant performance impact. The additional micro-batches to  peer workers and the dependencies between forward and backward pass computations lead to 9 additional time steps per iteration, for a total of 36. In other words, the failure of 8.3\% of workers (1 out of 12) leads to a 33\% slowdown in iteration time.

%While \emph{Adaptive Pipelines} enable resilient training, they can result in decreased training throughput under failures.
%This is because suddenly introducing extra micro-batches from a failed worker to the middle of a failure-free data pipeline can introduce pipeline bubbles as pipeline stages are stalled waiting for their respective inputs.
%Because all data parallel pipelines must synchronize each iteration, the entire training iteration latency is increased as a result.
%As shown in Figure~\ref{fig:fault-free-and-adaptive-schedule}, despite ensuring equal distribution of micro-batches across failure-free pipelines, \(W_{02}\) and \(W_{12}\) performed extra computation, resulting in pipeline bubbles which introduced a 9 time-step overhead compared to the fault-free schedule.

\subsection{Decoupled BackProp: Filling Unused Bubbles}
%\subsection{Decoupled Back Propagation}
\label{sec:decoupled-backprop}
\begin{figure}
    \centering
    \includegraphics[width=1\linewidth]{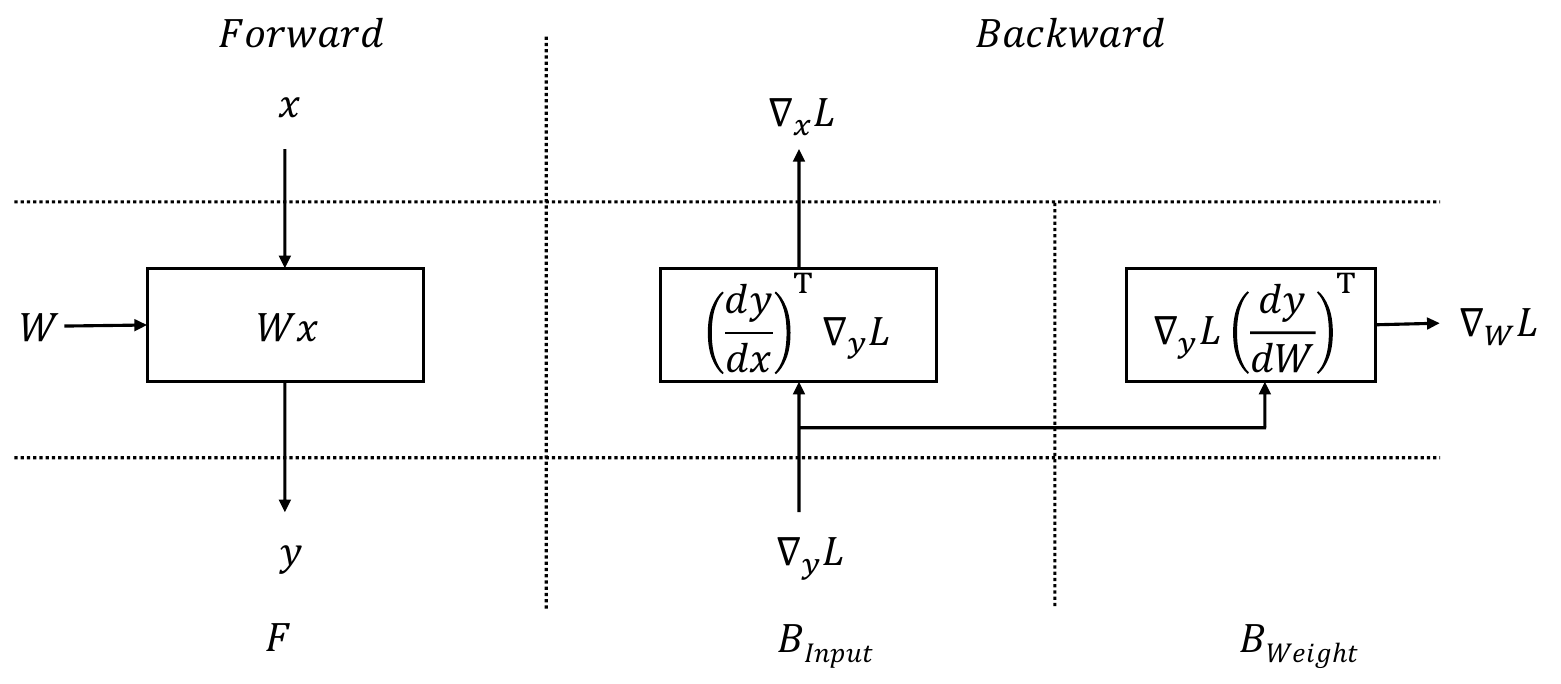}
    \caption{Forward and Backward pass for an operator.}
    \label{fig:split-backward}
\end{figure}
\begin{figure}
    \centering
    \begin{subfigure}[c]{\linewidth}
        \centering
        \includegraphics[height=1.5cm]{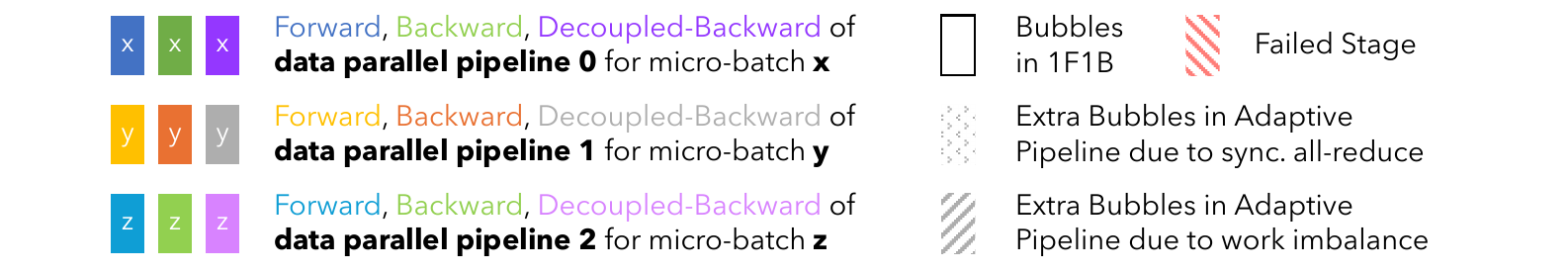}
    \end{subfigure}
    \hfill
    \begin{subfigure}[c]{\linewidth}
        \centering
        \includegraphics[width=\linewidth]{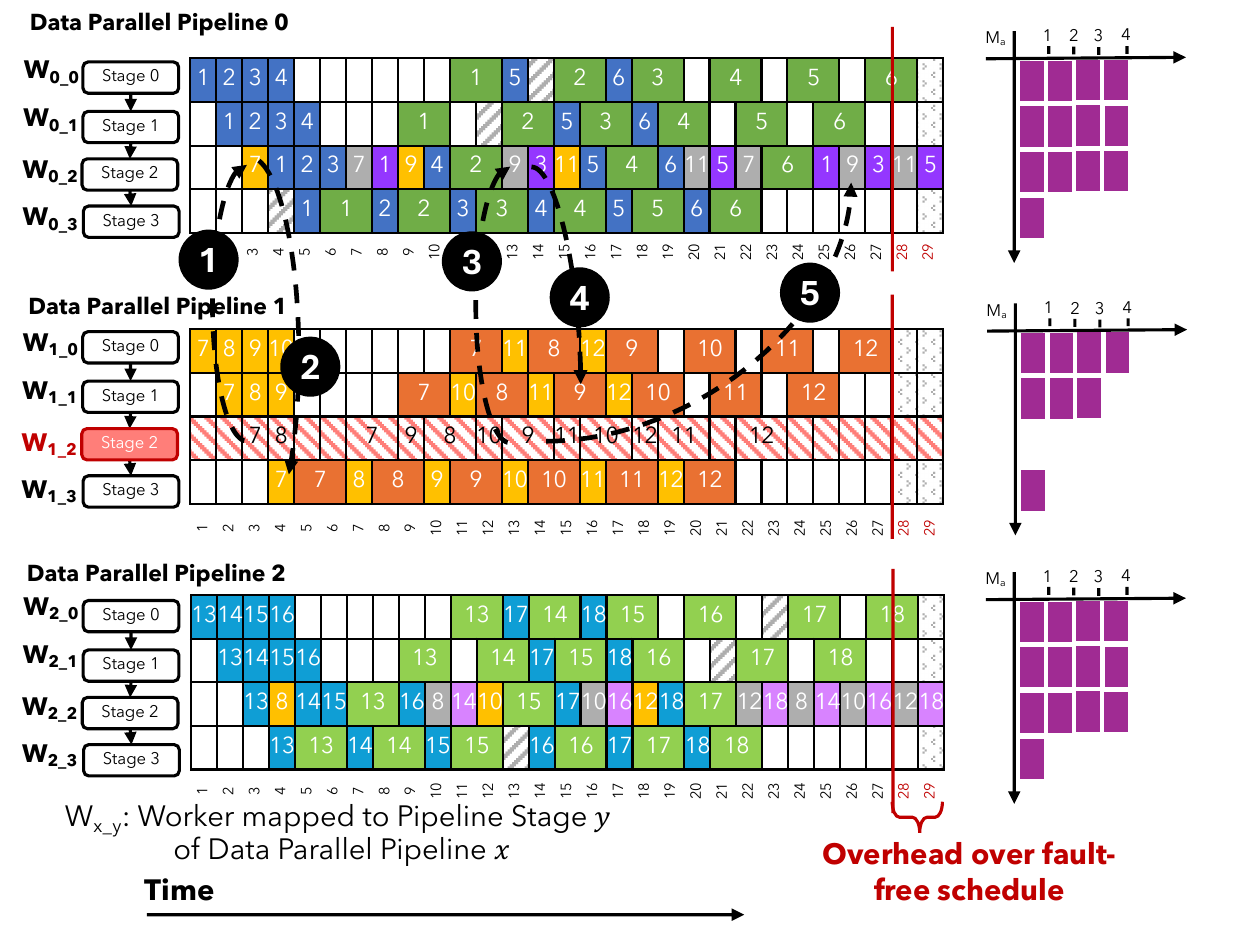}
    \end{subfigure}
    \caption{Optimized schedule with {\it Decoupled BackProp} when worker \(W_{1\textunderscore2}\) fails.}
    \label{fig:split-backward-adaptive-schedule}
    \vspace{-0.15in}
\end{figure}

\begin{figure*}[!t]
    \centering
    \includegraphics[width=1\textwidth]{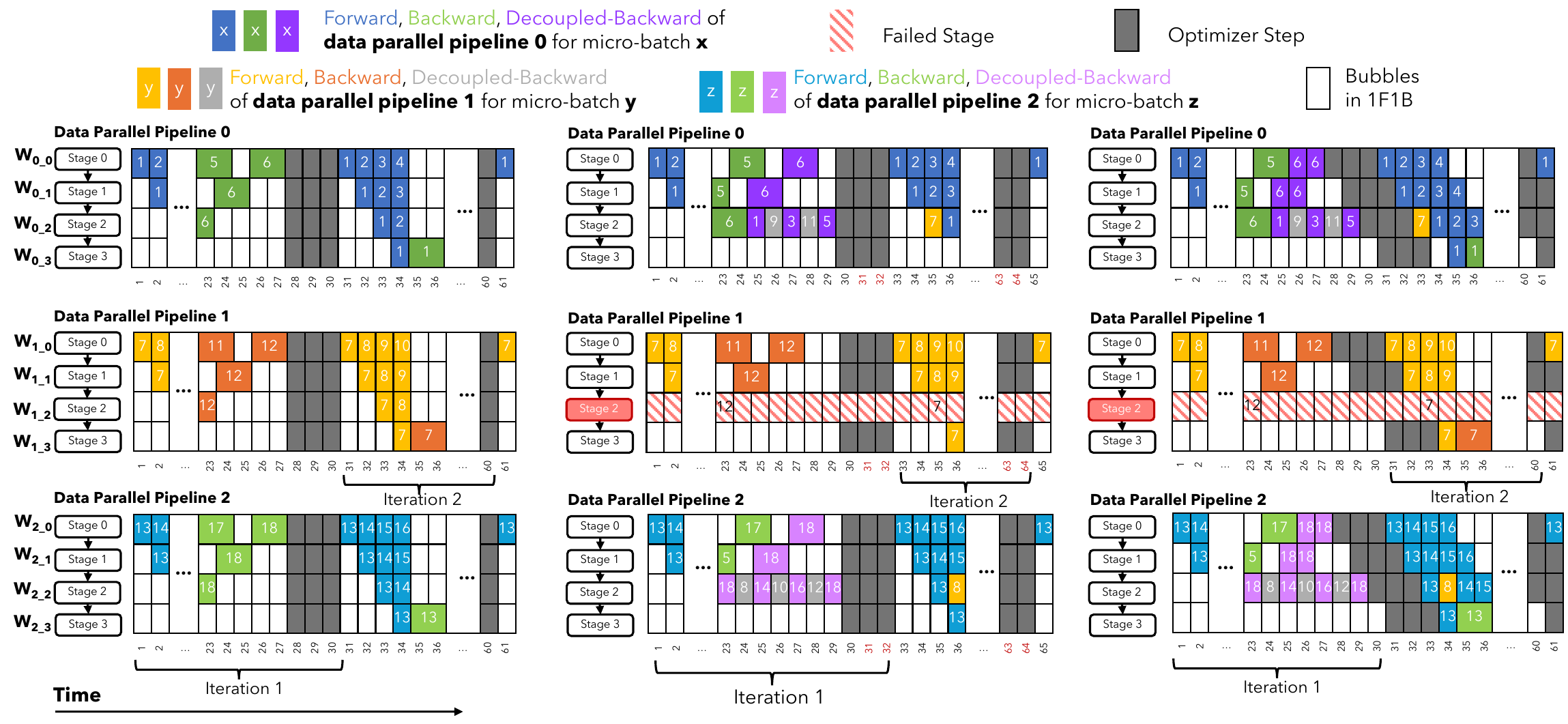}
    \begin{subfigure}[c]{0.3\linewidth}
        \centering
        \caption{Fault-free 1F1B Schedule}
    \end{subfigure}
    \hfill
    \begin{subfigure}[c]{0.32\linewidth}
        \centering
        \caption{Schedule for Adaptive Pipelining with Decoupled BackProp when worker \(W_{1\textunderscore2}\) fails.} % (can only use bubbles from cool down phase)}
        \label{fig:use-cool-down-bubbles}
    \end{subfigure}
    \hfill
    \begin{subfigure}[c]{0.32\linewidth}
        \centering
        \caption{Schedule with the addition of the Staggered Optimizer.} % (can use bubbles from warm up and cool down phase)}
        \label{fig:use-warm-up-cool-down-bubbles}
    \end{subfigure}
    \caption{The Staggered Optimizer allows \NAME to optimize pipeline schedule across training iterations.}
    %Comparative Scheduling Scenarios in \NAME with and without worker failure.}
    \label{fig:staggered-optimizer}
\end{figure*}

\textit{Decoupled BackProp} addresses the scheduling problem of \textit{Adaptive Pipelining}. It separates the backward pass into two distinct phases, allowing for more flexible scheduling of the extra load in the peers of a failed worker.
Figure~\ref{fig:split-backward} shows that the backward pass calculates two distinct outputs: the gradients relative to the input (\(B_{Input}\)) and the gradients relative to parameters (weights) for that pipeline stage (\(B_{Weight}\)).
% backpropagation calculates the calculates gradients relative to the inputs and the weights of an operator, which we denote as \(B_{Input}\) and \(B_{Weight}\), respectively.
Conventionally, \(B_{Input}\) and \(B_{Weight}\) are \emph{coupled} as a unified calculation.
%We make the critical observation that 
\textit{This coupling lengthens dependencies between pipeline stages.}
The backward pass for stage \(i\) must wait for the completion of both \(B_{Input}\) and \(B_{Weight}\) from stage \(i+1\), despite stage \(i\) requiring only \(B_{Input}\) for its backward computations. \(B_{Weight}\) \textit{can be deferred} to improve the overall pipeline schedule.
Thus, \textit{Decoupled BackProp} splits the backward pass into two distinct tasks.

Figure~\ref{fig:split-backward-adaptive-schedule} explains how \textit{Decoupled BackProp} reduces the overheads introduced by \textit{Adaptive Pipelining}. It shows the same example as in Figure~\ref{fig:fault-free-and-adaptive-schedule} with the addition of \textit{Decoupled BackProp}.
The forward pass for the micro-batch 7 is unchanged (\numcircledtikz{1} and \numcircledtikz{2}).
However, instead of waiting on a \emph{coupled} backward pass (both \(B_{Input}\) and \(B_{Weight}\)) from the previous stage to complete, which causes a ripple effect of dependencies throughout the pipeline, subsequent stages can continue processing new micro-batches using just the \(B_{Input}\) from their predecessors.
For example, worker \(W_{0\_2}\) can run the backward pass for micro-batch 9 (\numcircledtikz{3}) and immediately pass the \(B_{Input}\) to worker \(W_{1\_3}\) (\numcircledtikz{4}) without computing \(B_{Weight}\).
\(B_{Weight}\) is deferred to a later time (\numcircledtikz{5}) to reduce pipeline stalls, bringing the overall overhead down to just two time-steps (7.4\% overhead with 8.3\% failed workers).

Because \(B_{Weight}\) is dependence-free, it can largely be deferred until the end of the training iteration. Hence, we can take advantage of idle slots in the cool-down phase of the 1F1B schedule for \(B_{Weight}\) computations, freeing slots in the steady stage for the re-routed work of the failed worker. This enables \NAME to absorb the additional computational demands imposed by failures with minimal overheads.

%\paragraph{No free lunch: Why Decoupled BackProp must be used strategically.}
\noindent{\bf The memory challenge.}
Unfortunately, \textit{Decoupled BackProp} increases memory pressure. Decoupling the computation of \(B_{Weight}\) requires that intermediate data is stored for potentially extended periods on each worker. For instance, as shown in Figure~\ref{fig:split-backward-adaptive-schedule}, decoupling the backward pass for micro batch 1 necessitates retaining its intermediate data until time step 25.  %\christos{add one/two sentences that point to the memory part of figures 3 and 4} 
To avoid memory exhaustion, \NAME applies \textit{Decoupled BackProp} {\it selectively} only when it can mitigate the overheads of \textit{Adaptive Pipelining}. %, as we describe in  \S\ref{sec:planner}.
%In addition to being \emph{selective}, 
We also capitalize on the observation that memory imbalance exists among the pipeline stages~\cite{bpipe}. To ensure that all pipeline stages are fully utilized, earlier stages need to allocate additional memory to process more forward micro-batches than later stages~\cite{pipedream}. By exploiting this available \emph{surplus} memory, \NAME can effectively offset some of the memory demands of \textit{Decoupled BackProp} without incurring additional costs.

\subsection{Staggered Optimizer: Accessing More Bubbles}
%\subsection{Staggered Optimizer}
\label{sec:optimizer-validations}

\textit{Decoupled BackProp} makes efficient use of the cool-down bubbles. %to offset the extra work due to failed workers.
However, as seen in Figure~\ref{fig:staggered-optimizer}, the warm-up bubbles remain underutilized due to their placement \textit{after} the synchronous optimizer step of the previous iteration.
To exploit warm-up bubbles, we make the critical observation that optimizer steps for different pipeline stages are independent of each other.
The \textit{Staggered Optimizer} thus shifts the timing of the optimizer step across pipeline stages to better utilize the bubbles that occur during the warm-up phase of the subsequent training iteration.
Effectively, this staggering provides \NAME \textit{more} bubbles in the cool-down phase to hide the overhead of compensating for failed workers.

Figure~\ref{fig:use-warm-up-cool-down-bubbles} shows how combining \textit{Staggered Optimizer} with \textit{Adaptive Pipelining} and \textit{Decoupled BackProp} results in \textit{zero overhead} over the fault-free 1F1B schedule in the running example.
By staggering the optimizer step, later pipeline stages can move bubbles from the warm-up phase of the second iteration to the cool-down phase of the first iteration.
Thus, peers that need to process additional micro-batches from failed workers within their steady-state schedule (i.e, workers \(W_{0\_2}\) and \(W_{2\_2}\)) can further defer their weight gradient computations towards the cool-down phase. Workers for earlier stages (e.g., \(W_{0\_0}\) for stage 0) can continue with the optimizer step, ensuring that the entire pipeline does not stall. Worker \(W_{0\_0}\) can start iteration 2 at \textit{exactly the same time step as the fault-free 1F1B schedule}, even with failures.

Put together, these three mechanisms allow \NAME to flexibly optimize the use of computational resources across pipelines, ensuring that each stage operates near its capacity and maintains consistent utilization throughout training. 

\subsection{Supporting Multiple Failures and Re-Joins}\label{sec:multiple_failures}
\begin{figure}
    \centering
    \begin{subfigure}[c]{\linewidth}
        \centering
        \includegraphics[width=\linewidth]{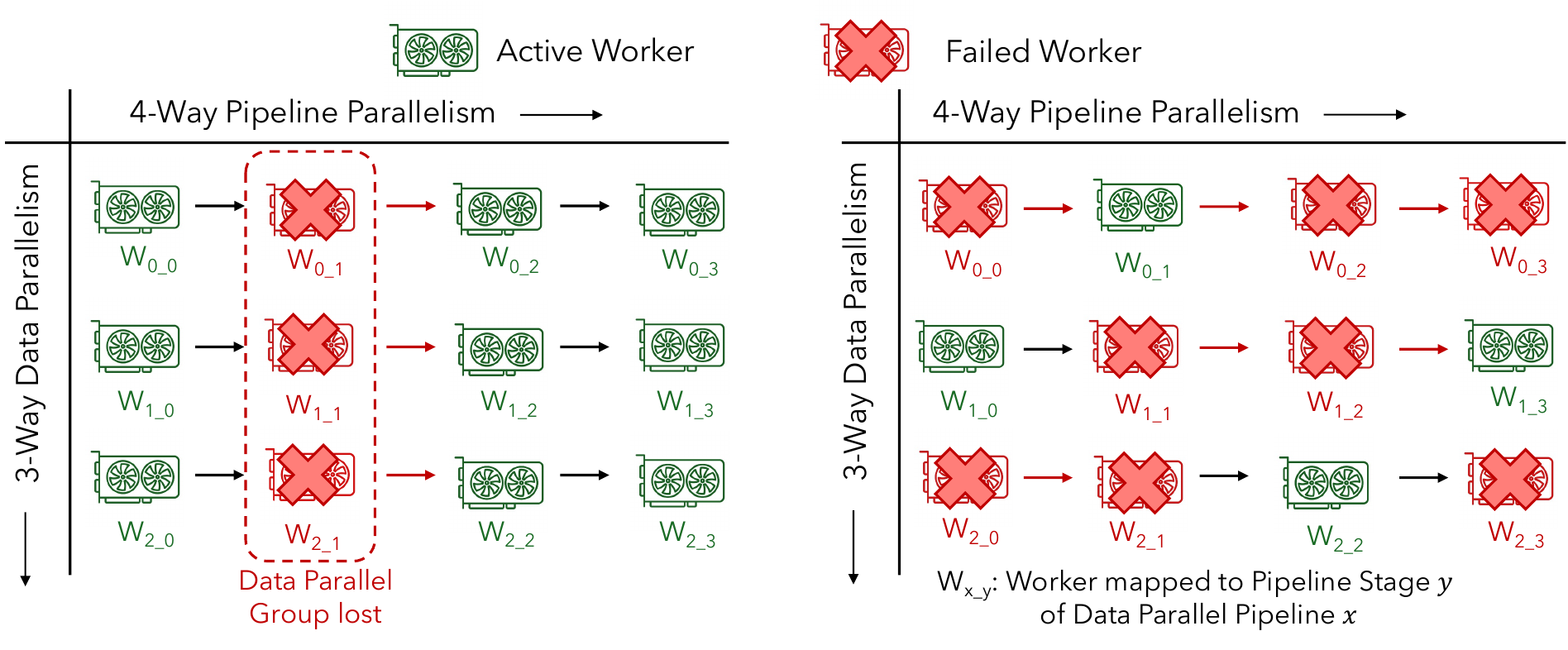}
    \end{subfigure}
    \hfill
    \centering
    \begin{subfigure}[c]{0.46\linewidth}
        \centering
        \caption{The loss of an entire DP group requires checkpoint restoration.}
        \label{fig:not-recoverable}
    \end{subfigure}
    \hfill
    \begin{subfigure}[c]{0.46\linewidth}
        \centering
        \caption{\NAME can continue training if a functional worker exists per stage.}
        \label{fig:recoverable}
    \end{subfigure}
    \caption{Example of \NAME's fault tolerance guarantees.
    % \textbf{(a)} The loss of an entire DP group requires checkpoint restoration.
    % \textbf{(b)} \NAME can continue training if a functional worker exists per stage.
    } % with data parallelism (DP) degree of 3.}
    \label{fig:recovery-in-recycle}
\end{figure}

%Our example, thus far, considered a single worker failure. 
\NAME supports multiple failures and rejoins. %leverages the same mechanisms to resume execution from multiple worker failures, rerouting the work of each failed worker to its data parallel peers.
\NAME guarantees continued training up to \(DP-1\) simultaneous worker failures, as in the worst case \(DP\) failures can disable all the data-parallel peers of a pipeline stage.
%because in the worst \(DP\) failures are enough for all data-parallel peers in a pipeline stage to fail simultaneously.
% data parallelism degree.
%For the example in Figure~\ref{fig:recovery-in-recycle}, where $DP=3$, \NAME can withstand up to 2 simultaneous worker failures.
However, \NAME can probabilistically sustain more failures, as long as there is at least one functional worker per stage across all data parallel pipelines. 
Figure~\ref{fig:recoverable} shows how \NAME can recover from more than \(DP-1\) failures. Despite 8 simultaneous failures -- \textit{2/3 of the GPUs!} -- \NAME ensures continual training. %at least one worker in each stage is still operational, allowing \NAME to continue training. 

As workers are repaired, \NAME can re-insert them back into the adapted pipeline schedule, reducing the probability of high counts of concurrent failures. Unlike unexpected failures, planned worker additions occur at iteration boundaries to overlap any data copying overhead with the previous iteration. Once a worker rejoins, \NAME ceases rerouting micro-batches and sends all micro-batches from the previous stage to the recovered worker.

 In the worst-case scenario, where simultaneous failures impact a single data-parallel group (as shown in Figure~\ref{fig:not-recoverable}), \NAME falls back on existing resilience strategies: it calculates an efficient hybrid-parallel scheme for the remaining nodes, configures it across all workers by restoring from a recent checkpoint, and resumes training at the speed supported by the new parallelism.

%As the number of concurrent  failures increases, it becomes more difficult to schedule the additional work on peer workers without extending the 1F1B schedule. In the example used in this section, \NAME recovers with zero performance overheads. As long as we keep the performance loss from $x\%$ failures below or close to $x\%$, \NAME is efficient. As we will show in \S\ref{sec:evaluation}, \NAME is effective in optimizing performance under failures for a wide range of scenarios.

\noindent{\bf Multi-GPU Servers and Tensor Parallelism.} Training systems commonly employ servers equipped with multiple GPUs.
A favoured configuration features HGX/DGX servers with 8 GPUs connected in an all-to-all manner using NVLink and NVSwitches~\cite{dgx}. Similar to Oobleck and Bamboo, we treat an entire server as the unit of failure. This approach is pragmatic because servicing even one faulty GPU board necessitates taking the entire server offline. Additionally, issues such as software malfunctions, CPU issues, connectivity disruptions, and power or cooling issues can incapacitate an entire server. Most training systems implement tensor parallelism within a multi-GPU server to leverage the high-bandwidth NVLink connections. However, a tensor parallel group can extend across multiple servers. In such cases, \NAME, along with Oobleck and Bamboo, classifies the entire group of tensor-parallel servers as a single unit of failure.

\section{\NAME Design}\label{sec:design}

%\NAME realizes the benefits of \textit{Adaptive Pipelining}, \textit{Decoupled BackProp}, and the \textit{Staggered Optimizer} in an end-to-end training system.
%\myzhao{If we have space, a figure that shows these components will be useful}
%As shown in Figure XX, \NAME consists of three critical components, a \textit{Profiler}, \textit{Executor}, and critically a \textit{Planner}.
%We begin by walking through how \NAME leverages these components to manage and ensure the resiliency of the training job.
%Section~\ref{sec:planner} then describes how the \textit{Planner} leverages the key techniques to generate efficient training schedules even under failures.

\subsection{System Overview}\label{sec:design_overview}

\begin{figure}
    \centering
    \includegraphics[width=1\linewidth]{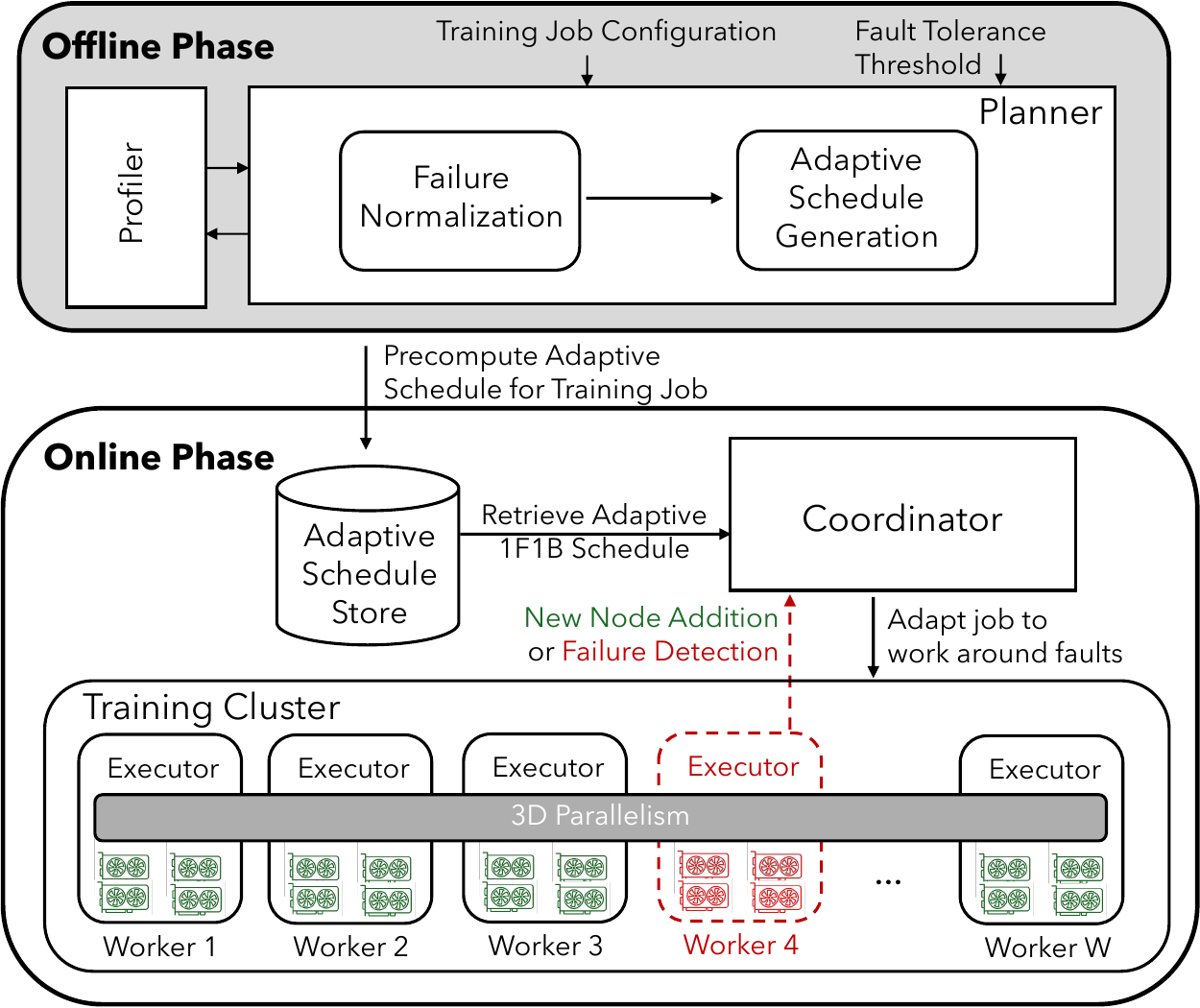}
    \caption{\NAME system overview.}
    \label{fig:system-overview}
\end{figure}

\NAME consists of three key components, the \textit{Profiler}, the \textit{Executors}, and critically the \textit{Planner}, as shown in Figure~\ref{fig:system-overview}.

\noindent\textbf{Profiler.}
When a large training job is first submitted, 
\NAME runs a short profiling job to collect key performance statistics such as the average micro-batch latency for the forward and backward passes, memory requirements for activations and gradients, and inter-node communication bandwidth. 
The profiling job executes a small number of training iterations, 100 by default, and typically takes a few minutes. 
These statistics are used by the \textit{Planner}.
%The \textit{Profiler} aggregates the collected statistics for use by the \textit{Planner}.

\noindent\textbf{Planner.} The \textit{Planner} generates pipeline schedules for the training job under various failure scenarios. %\christos{check} \myzhao{So the "near" optimal is introduced b/c we can use a heuristic cost model in the DP, otherwise the DP/ILP should be optimal. I guess "near optimal" is the best way to say it?...} 
It uses dynamic programming and mixed integer linear programming (MILP) to implement the \NAME techniques (Section~\ref{sec:system}). 
We generate a pipeline schedule for each \textit{number} of tolerated simultaneous failures. Each plan is agnostic to which specific worker(s) fail. It encodes the specific pipeline schedule, specifying micro-batch assignments, the sequence of forward and backward tasks, and communication of activations and gradients.
The plans are stored in distributed fault-tolerant storage (e.g., \textit{etcd}) to be used by runtime \textit{Executors}. % during runtime.

By default, we generate up to $DP-1$ plans, which accounts for the $DP-1$ simultaneous failures that \NAME can always handle. Since \NAME can probabilistically sustain many more failures, we can also generate plans for up to a user-defined fault tolerance threshold. 
%The \textit{Planner} runs relatively quickly, requiring 53 minutes to generate plans for up to 512 concurrent failures for a cluster of 2048 GPUs. Nevertheless, the training job can begin training in parallel as most plans will not be needed immediately.

%the training job can begin training in parallel as most plans will not be needed immediately. 
%; ${\sim}50$ minutes to generate plans for up to 512 concurrent failures for a cluster of 2048 GPUs.
%Nevertheless, the training job can begin training in parallel as most plans will not be needed immediately. 

\noindent\textbf{Executors.} \NAME operates with an \textit{Executor} on each GPU node, tasked with implementing the training plan specific to that node. Overseeing these Executors, a centralized \textit{Coordinator} orchestrates the overall training operation by distributing the appropriate execution plans to each Executor. Additionally, the \textit{Coordinator} actively monitors the status of all Executors to ensure their continuous operation. In the event of a failure, the \textit{Coordinator} reassigns the tasks impacted by failure to the most suitable location, as advised by the \textit{Planner} (refer to Section~\ref{sec:normalization-phase}), and then directs the remaining \textit{Executors} to resume training according to the updated plan that now reflects the updated count of non-operational workers. Training resumes from the iteration during which the failure was identified. If failures are not promptly detected, restoring a checkpoint from remote storage may become necessary. To enhance reliability, the \textit{Coordinator} itself is safeguarded against failures through active replication.

\subsection{Planner}\label{sec:planner}
Given the cluster configuration, training job, profiling statistics, and failed workers, the \textit{Planner} calculates an adaptive schedule that minimizes the training iteration latency using the fault-free workers. It does this in two phases, \textit{Failure Normalization} followed by \textit{Adaptive Schedule Generation}.

To avoid solving an MILP for the combinatorial number of possible failure locations, the \textit{Failure Normalization} phase first normalizes each given \textit{number} of failures by calculating the suitable location within the pipeline schedule to migrate each failure.
%\NAME generates an execution plan for a given \textit{number} of failures, regardless of \textit{which} worker(s) fails, as solving ILP problems for the combinatorial number of possible failure locations is intractable.
%\emph{Failure Normalization} first \textit{normalizes} the specific failures by migrating failed nodes to a pre-determined and ideal set of locations within the pipeline schedule.
%Thus, regardless of the initial worker failure locations, \NAME first re-orders the 3D parallelism groups to ensure that the second (and expensive) phase only needs to generate one solution for a given number of failures.
For example, if \textit{any} single node fails in the example in Figure~\ref{fig:recovery-in-recycle}, the {\it Planner} will swap the node with the calculated ideal location, e.g., $W_{2\_3}$.
Unlike current solutions which require a complete reconfiguration of a data-parallel pipeline~\cite{oobleck}, normalization only requires a point-to-point copy of model parameters to swap the location of two workers in the pipeline for each failure.
The \textit{Adaptive Schedule Generation} then leverages a MILP to derive an adaptive pipeline schedule for each normalized case. 

%\christos{maybe 1 sentence on how to project back to the specific failure case? the naive approach requires a lot of parameter re-shuffling which we claim we don't do much}

\subsubsection{Failure Normalization}\label{sec:normalization-phase}

\begin{algorithm}[h]
\caption{Failure Normalization}\label{alg:normalization}
\begin{algorithmic}[1]
\State $DP \gets$ Number of data-parallel pipelines.
\State $PP \gets$ Number of pipeline stages.
\State $MB \gets$ Number of microbatches per pipeline.
\State $F \gets$ Total number of failures.
%\Comment $DP[i][j]$ represents the cost of handling $j$ failures across pipeline stages $[0, i]$
\State $O \gets$ an PP $\times$ (F+1) array \Comment Rerouting Overheads
\State $A \gets$ an PP $\times$ (F+1) array  \Comment Assignments
%[i][0] \gets 0 \quad \forall i \in \{0, \dots, N-1\}$
%\State $M[i][0] \gets [] \quad \forall i \in \{0, \dots, N-1\}$
\State \Procedure{failure\_reordering}{$DP, PP, MB, F$}
\For{$i \in \{0, \dots, PP-1\}$}
\For{$f \in \{0, \dots, F\}$}
\If{$i == 0$}
\State $O[i][f] = \Call{COST}f$
\State $A[i][f] = [f]$
\Else
\State $x = \argmin_{x \leq f}\Bigl(O[i-1][f-x] $  \\ \hfill $ + ~\Call{COST}x\Bigr)$
\State $O[i][f] = O[i-1][f-x] + \Call{COST}x$
\State $A[i][f] = concat\Bigl(A[i-1][f-x], x\Bigr)$
\EndIf
\EndFor
\EndFor \\
\Return $A[PP-1][F]$
\EndProcedure

\State \Procedure{cost}{$f$}
 \If{$f > 0$} \State
 \Return $\min\Bigl(0, MB \times f \times 3 - (DP - f) \times (PP - 1) \times 3\Bigr)$
 \EndIf
 \State \Return $0$
\EndProcedure
\end{algorithmic}
\end{algorithm}

The intuition behind \textit{Failure Normalization}, shown in Algorithm~\ref{alg:normalization}, is twofold: a) distribute failures across different peer groups to enhance fault tolerance and evenly balance the additional workload, and b) shift failures to later pipeline stages with more bubbles to maximize the opportunity to hide re-routing overheads.
%(i.e., to later stages).
%Distributing failures across peer groups maximizes fault tolerance and load balances extra work across different peer groups.
%Increasing the number of bubbles maximizes the opportunity to hide overheads.% \NAME techniques can introduce.

\textit{Failure Normalization} uses dynamic programming to compute a migration strategy for handling a given number of failures. Given a total of $F$ failures and $PP$ pipeline stages, it returns a list $A$ of length $PP$, where the sum of elements in $A$ equals $F$. Each $A[i]$ specifies the number of failures assigned to pipeline stage $i$ across all data parallel pipelines, with the specific pipeline assignments being arbitrary and not impacting performance. For example, if $A[3]=2$, two failures would be assigned to stage 3, and we could swap the failed nodes with any two from that stage, such as $W_{0\_3}$ and $W_{2\_3}$ in Figure~\ref{fig:recoverable}.

The overhead of handling $f$ failures at stages from $0$ to $i$ is represented by $O[i][f]$, while $A[i][f]$ indicates how the $f$ failures are distributed across these stages, with $A[i][f]$ being of length $i+1$. The recurrence relation minimizes overhead by finding the optimal failure assignment for $f$ failures across the first $i$ stages, and the best assignment for $F$ failures is given by $A[PP-1][F]$.

To compute $COST(i, x)$, we estimate the additional time slots needed for handling extra micro-batches due to $x$ failures in the $i$-th stage using the heuristic from line 27. While the MILP in Section~\ref{sec:solution} could be used, \textit{Failure Normalization} opts for the heuristic to reduce computation time. The complexity of determining the migration strategy for $F$ failures and $PP$ pipeline stages is $O(PP \times F)$.

%We defer details of the DP formulation to the Appendix.

%Thus, upon any detected failure, the \textit{Coordinator} will first perform the migration strategy to normalize the failed nodes according to the DP before proceeding with the optimized execution plan.
%\myzhao{@swapnil, i'm assuming this is what would need to happen? If so, what explicitly would this do? shuffle weights to necessary workers? anything else?}
%\myzhao{I believe that we need a copy of parameters to migrate the failed worker to the right place in the pipeline. Do we want to say anything about that?}

\subsubsection{Adaptive Schedule Generation}\label{sec:solution}

Next, the \textit{Planner} uses normalized failure locations and profiled statistics as inputs for an MILP to generate adaptive schedule. At a high level, the MILP determines how to re-route micro-batches across peers of a failed worker, integrating both regular and re-routed micro-batches (\textit{Adaptive Pipelining}). It considers communication latency $T_{comm}$, micro-batch latency for forward, backward input and weight pass -- $T_{F}$, $T_{B_{Input}}$ and $T_{B_{Weight}}$ respectively, task dependencies in forward and backward passes, and memory usage, leveraging the \textit{Decoupled BackProp} and \textit{Staggered Optimizer} techniques.

\noindent\textbf{Notation.} Each operation in a training iteration is denoted by the 5-tuple $(i, j, k, c, k_s)$. Here, $i$ represents the pipeline stage within each data parallel pipeline, and $k$ indicates the operation's original data parallel pipeline before any failures. The variable $j$ denotes the micro-batch ID in the training iteration, and $c$ specifies the type of operation for each micro-batch, where $c \in \{F, B_{input}, B_{weight}\}$. Finally, $k_s$ specifies the peer pipeline that executes a micro-batch originally intended for $k$; which can be the same as $k$. For instance, a micro-batch ID $14$ originally scheduled for $W_{2\_3}$ but rerouted to peer $W_{1\_3}$ would be identified with $i=3$, $j=14$, $k=2$, and $k_s=1$.

\noindent\textbf{Inputs.}
We use micro-batch assignments for each worker as inputs. The \textit{Planner} reassigns micro-batches from failed workers to peer workers within the same group while retaining the original micro-batches. This results in a binary mapping $S_{i,j,k}^{k_s} \in \{0, 1\}$, indicating whether a micro-batch $(i, j, k)$ should run on pipeline $k_s$. Each operation is assigned to exactly one pipeline: \( \sum_{k_s} S_{i,j,k}^{k_s} = 1 \).

Additionally, we use profiled statistics, which include $T_c$ for the computational time of each operation $c$, $T_{comm}$ for communication latency of activations or gradients\footnote{We use a single value as activations and gradients are the same size.}, and $\Delta M_{i,j,k,c}^{k_s}$ for the \textit{change} in memory utilization on worker $(i, k, k_s)$ due to the execution of operation $(i, j, k, c, k_s)$.

\begin{equation*}
\Delta M_{i,j,k,c}^{k_s} =
\begin{cases}
    A_{B} & \text{, if } c=F \text{ and  } S_{i,j,k}^{k_s} = 1 \\
    A_B - A_{B_{Input}} & \text{, if } c=B_{Input} \text{ and } S_{i,j,k}^{k_s} = 1 \\
    -A_{B_{Weight}} & \text{, if } c=B_{Weight} \text{ and } S_{i,j,k}^{k_s} = 1 \\
    0 & \text{, otherwise}
\end{cases}
\end{equation*}

Here, \(A_B\) is the profiled size of activation at the end of the forward pass. \(A_{B_{input}}\) and \(A_{B_{weight}}\) are the sizes of gradients at the end of the backward-input and backward-weight passes, respectively. We free \(A_{B_{input}}\) and \(A_{B_{weight}}\) after completing their respective backward passes.

\sloppy
\noindent\textbf{Variables.}
We define a binary variable \(O_{(i,j,k,c,k_s) \rightarrow (i',j',k',c',k_s')} \in \{0, 1\}\) to represent ordering between pair of operations \((i, j, k, c, k_s)\) and \((i', j', k', c', k_s')\). This variable is 1 if \((i', j', k', c', k_s')\) is scheduled after \((i, j, k, c, k_s)\), and 0 otherwise. Ordering is needed only within a stage and between computation phases of the same micro-batch (e.g., $F$ and $B_{input}$). Additionally, \(E_{i,j,k,c}^{k_s}\) represents the ending time of operation \((i, j, k, c, k_s)\).

\noindent\textbf{Objective.}\label{sec:appendix_milp}
Our objective is to minimize the makespan of a single training iteration while adhering to task dependencies and memory constraints. This involves determining the sequence and timing of operations throughout the training pipeline using the sets of variables $O$ and $E$. Thus, the objective can be formulated as follows:

\begin{equation}
    \min_{\substack{O, E}} \qquad \max_{\substack{i,j,k,k_s}} E_{i,j,k,B_{Weight}}^{k_s} 
\label{eq:ilp-optim-target}
\end{equation}

Here, $\max_{i,j,k,k_s} E_{i,j,k,B_{Weight}}^{k_s}$ represents the end time of the \textit{last} operation in the iteration, representing the makespan.

\noindent\textbf{Constraints.}
The constraints applied to the objective are the following.

\noindent\underline{\textit{Cross-Stage Dependencies.}}
\begin{equation}
E_{i,j,k,F}^{k_s} \ge 
    S_{i,j,k}^{k_s} \times 
        (\sum_{\hat{k}} (E_{i-1,j,k,F}^{\hat{k}} \times S_{i-1,j,k}^{\hat{k}}) \\ + T_{comm} + T_{F})
\label{eq:ilp-cross-stage-dep-forward}
\end{equation}

\begin{equation}
    \begin{multlined}
    E_{i,j,k,B_{Input}}^{k_s} \ge 
        S_{i,j,k}^{k_s} \times
            (\sum_{\hat{k}} (E_{i+1,j,k,B_{Input}}^{\hat{k}} \times S_{i+1,j,k}^{\hat{k}}) \\ + T_{comm} + T_{B_{Input}})
    \end{multlined}
\label{eq:ilp-cross-stage-dep-backward}
\end{equation}

Equation~\ref{eq:ilp-cross-stage-dep-forward} specifies the dependency of a given micro-batch's forward pass on previous pipeline stages (e.g., stage 0 must execute before stage 1).
Similarly, Equation~\ref{eq:ilp-cross-stage-dep-backward} specifies the reverse dependency for the backward pass (e.g., stage 1 must execute before stage 0).

\noindent\underline{\textit{Same-Stage Dependencies.}}
\begin{equation}
E_{i,j,k,B_{Weight}}^{k_s} \ge 
    S_{i,j,k}^{k_s} \times (E_{i,j,k,B_{Input}}^{k_s} + T_{B_{Weight}})
\label{eq:ilp-same-stage-dep}
\end{equation}

Equation~\ref{eq:ilp-same-stage-dep} allows the MILP to reason about \textit{Decoupled BackProp}, specifying that $B_{input}$ must precede $B_{weight}$ for a given micro-batch within a worker.

\noindent\underline{\textit{No Overlapping Computations.}}
\begin{equation}
    \begin{multlined}
    E_{i,j^\prime,k^{\prime},c^\prime}^{k_s^{\prime}} \ge 
    E_{i,j,k,c}^{k_s^{\prime}} + T_{c^\prime} - \\ \infty (1 - S_{i,j,k}^{k_s^{\prime}} \times S_{i,j^\prime,k^\prime}^{k_s^{\prime}} + O_{(i,j,k,c,k_s^{\prime})\rightarrow(i,j^\prime,k^\prime,c^\prime,k_s^{\prime})})
    \end{multlined}
\label{eq:ilp-no-overlap}
\end{equation}

Equation~\ref{eq:ilp-no-overlap} adds a dependency constraint which specifies that two different operations cannot overlap in time if they are executed on the same worker.
For example, if operation $(i, j', k', c', k_s^\prime)$ is scheduled after $(i, j, k, c, k_s^\prime)$, they both execute on worker $(i, k_s^\prime)$ and thus $(i, j', k', c', k_s^\prime)$  must begin after $(i, j, k, c, k_s^\prime)$ ends.

\noindent\underline{\textit{Memory Constraint.}}
\begin{equation}
    \begin{multlined}
    M_{Limit} \ge \Delta M_{i,j^\prime,k^\prime,c^\prime}^{k_s^{\prime}} + \\ \sum_{j,k,c} \Delta M_{i,j,k,c}^{k_s^{\prime}} \times O_{(i,j,k,c,k_s^{\prime})\rightarrow(i,j^\prime,k^\prime,c^\prime,k_s^{\prime})}
    \end{multlined}
\label{eq:ilp-memory}
\end{equation}

Finally, Equation~\ref{eq:ilp-memory} calculates the activation memory required at any given time on worker $(i, k_s^\prime)$, and constrains operations such that the total memory is below an $M_{limit}$.

\section{Implementation}\label{sec:implementation}
We implemented \NAME on top of DeepSpeed~\cite{deepspeed} and made the following additions to DeepSpeed to support \NAME. 

\noindent {\bf Detecting failures.} \NAME uses existing hardware error detection mechanisms in GPUs (e.g.; SMBPBI APIs in NVIDIA GPUs) and software error detection in runtime systems (e.g. stderr logs in PyTorch). Workers send periodic heartbeats to a central driver, including worker information and hardware statistics from GPUs such as page retirement count, row-remapping stats, ECC correction stats, XID error logs, and stdout/stderr logs~\cite{megascale}. If the central driver detects abnormalities, it marks the worker as failed.

\noindent {\bf Rerouting micro-batches to data parallel peers.}  To handle the dynamic rerouting of micro-batches following node failures, we introduce two new and complementary communication operators: \texttt{ReRouteAct} and \texttt{ReRouteGrad}. Positioned at the end of a pipeline stage, \texttt{ReRouteAct} normally acts as a pass-through, transmitting computed intermediate activations to the next stage within the same data parallel pipeline. Conversely, \texttt{ReRouteGrad}, located at the beginning of a stage, typically forwards gradients backward through the pipeline. When a subsequent stage fails, \texttt{ReRouteAct} redistributes the micro-batches across the remaining peers in a round-robin fashion, while \texttt{ReRouteGrad} adjusts the gradient distribution to ensure that both the forward and backward processes of a micro-batch are handled by the same peer. This strategy maintains operational continuity in \NAME under fault conditions, mirroring fault-free execution semantics. These operators are integrated as \emph{pipeline instructions} within the DeepSpeed execution engine.

%These complementary operators playing a critical role in dynamically managing the distribution and aggregation of intermediate activations and gradients among the peers within a data-parallel group during disruptions.

\noindent {\bf Decoupling Back Propagation in DeepSpeed.}
Our implementation of \emph{Decoupled BackProp} within DeepSpeed centers around intercepting the weight gradient computations traditionally performed during the backpropagation phase. These computations are temporarily held in a newly created in-memory structure known as \emph{WeightGradStore}, enabling \NAME to defer weight gradient computation. To facilitate this, two new \emph{pipeline instructions} have been introduced into DeepSpeed's execution engine: \texttt{InputBackwardPass} and \texttt{WeightBackwardPass}. These instructions are incorporated into \NAME's execution plan and are then processed by DeepSpeed’s execution scheduler, effectively managing the distinct phases of backpropagation.

\noindent {\bf Bypassing Optimizer Synchronizations.} In DeepSpeed, each training iteration ends with an all-reduce collective to synchronize gradients across data-parallel peers, followed by validation checks at each pipeline stage to ensure numerical stability. If any stage detects a potential issue, the optimizer step for that iteration is skipped to prevent further problems. Upon successful validation, the optimizer step is executed, allowing synchronized progression to the next iteration.

\begin{table*}[!t]
  \captionsetup{justification=centering}
  \caption{Training throughput (samples/sec) with increasing failure frequency, higher is better.
  %Fraction of fault-free throughput (higher is better) achieved across increasing failure rates. 
  Bamboo ran out of memory for GPT-3 3.35B and 6.7B.}\label{tbl:throughput-comparision}
  \centering
  %% \small
  \begin{tabular}{c|ccc|ccc|ccc}
  \toprule
    \textbf{Systems} & \multicolumn{3}{c|}{\textbf{GPT-3 Medium}} & \multicolumn{3}{c|}{\textbf{GPT-3 3.35B}} & \multicolumn{3}{c}{\textbf{GPT-3 6.7B}} \\  \cline{1-1}
    \textbf{Failure Frequency} & \textbf{6h} & \textbf{2h} & \textbf{30m} & \textbf{6h} & \textbf{2h} & \textbf{30m} & \textbf{6h} & \textbf{2h} & \textbf{30m}  \\ \hline
   Fault-Free DeepSpeed~\cite{deepspeed} & \multicolumn{3}{c|}{27.58} & \multicolumn{3}{c|}{14.87} & \multicolumn{3}{c}{5.33} \\
   Bamboo~\cite{bamboo} & 19.47& 18.98& 15.24& OOM& OOM& OOM& OOM& OOM& OOM \\
   Oobleck~\cite{oobleck} & 27.26& 25.37& 19.47& 14.55& 13.44& 9.78& 4.98& 4.65& 2.78 \\
   \NAME & 27.27& 25.42& 22.27& 14.59& 14.17& 12.63& 5.17& 4.85& 3.53 \\
   \bottomrule
  \end{tabular}
\end{table*}

In contrast, \NAME uses a staggered timing approach for optimizer steps across different pipeline stages to improve scheduling efficiency. This method of staggering is not conducive to cross-stage synchronization for validating numerical stability prior to executing the optimizer. To address this, \NAME shifts numerical validations from a pre-step to a post-step process. After the all-reduce collective, each stage performs its own local validation checks without waiting for downstream stages. Each stage executes its optimizer step based on its own validation results and those of preceding stages. If a downstream stage fails validation, a rollback occurs across all stages before moving to the next iteration. Notably, for many optimizers, including the commonly used AdamW~\cite{adam}, these rollbacks incur no additional memory costs due to the arithmetic reversibility of the operations. This adjustment enables \NAME to leverage staggered operations while safeguarding the training process from potential instabilities.

\section{Evaluation}\label{sec:evaluation}

\begin{figure*}
    \centering
    \includegraphics[width=1\linewidth]{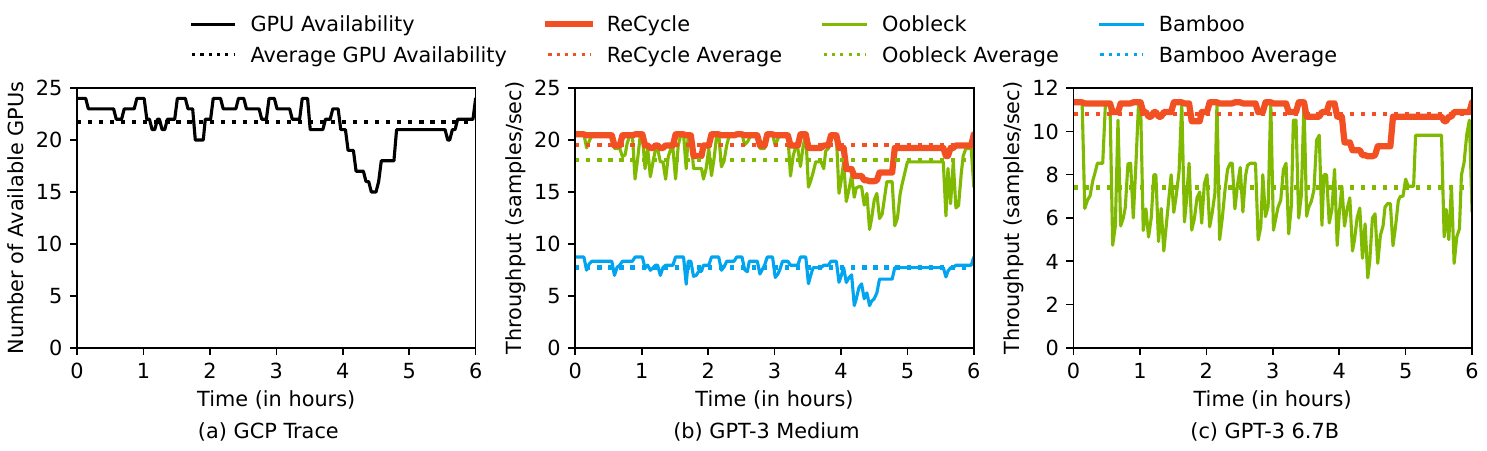}
    \caption{Training throughput (samples/sec), higher is better, for the GPT-3 Medium and GPT-3 6.7B models over the GCP trace. In \ref{fig:real-world-trace}b and \ref{fig:real-world-trace}c, the dashed lines represent the average training throughput achieved by each system within the 6h period.}
    \label{fig:real-world-trace}
\end{figure*}

\subsection{Experimental Setup}\label{sec:exp_setup}
\noindent\textbf{Cluster Setup.} We conducted real-world experiments on a 32-GPU cluster featuring NVIDIA A100 80GB GPUs in Azure, utilizing Standard\_NC96ads\_A100\_v4 instances (8 GPUs, 96 vCPUs, and 880 GB memory each). Each node includes a 600 GB/s NVLink intra-node interconnect and a 640 Gbps inter-node interconnect across 8 NICs. Scalability experiments were conducted using a simulator, discussed in Section~\ref{sec:exp_scalability}.

\noindent\textbf{Baselines.} 
We evaluated \NAME against two state-of-the-art baselines discussed in Section~\ref{sec:background_systems} -- Bamboo~\cite{bamboo} and Oobleck~\cite{oobleck}. Both baselines were tested with all their optimizations enabled.
Additionally, we report the \textit{fault-free} throughput achieved by DeepSpeed~\cite{deepspeed} using a 1F1B pipeline schedule~\cite{pipedream}. All experiments were conducted on the same Azure cluster for consistency.

\noindent\textbf{Workloads.} We evaluated all systems using the Megatron~\cite{megatron-lm} implementation of GPT-3~\cite{gpt-3}, with three model sizes: Medium (350M), 3.35B, and 6.7B. The applied (PP, DP) degrees were (2, 16), (4, 8), and (8, 4), respectively, with a TP degree of 1 for all models. Training was conducted on wikitext~\cite{wikitext} with batch and micro-batch sizes of (8192, 8) for Medium, and (1024, 1) for both 3.35B and 6.7B models. All real-world experiments, unless stated otherwise, ran for 6 hours with 32 workers.

\subsection{Training Throughput Under Failures}\label{sec:exp_degradation}

\paragraph{How well does \NAME handle failures compared to baselines?}
We first evaluate the average training throughput of Bamboo, Oobleck and \NAME on various failure scenarios. We set the frequency of failures from once every 6 hours to once every 30 minutes to cover a wide spectrum of environments~\cite{microsoft-training-trace, mlaas-in-the-wild}. We monotonically reduce the number of available workers without recovery, so the total number of workers steadily decreases over the course of each experiment -- for example in the 30m case, only $62.5\%$ of the workers (20 out of 32 workers) remain at the end of training.

Table~\ref{tbl:throughput-comparision} presents the average throughput for various failure frequencies and model sizes. Bamboo suffers from static overhead due to redundant computations and additional model state copies, which quickly depletes GPU memory. This leads to its inability to train larger models and a significant drop in throughput for even the smallest model—resulting in a 29\% reduction in throughput in the 6-hour failure frequency scenario. In contrast, Oobleck effectively manages all model sizes and demonstrates throughput improvements over Bamboo. However, Oobleck's performance declines with increasing failure frequency and model size due to imbalanced heterogeneous pipelines and higher reconfiguration latency.

\NAME effectively continues training despite failures, consistently matching or exceeding the throughput of both baselines across all models and failure rates. Unlike Bamboo, the techniques utilized in ReCycle—{\it Adaptive Pipelining}, {\it Decoupled BackProp}, and {\it Staggered Optimizer}—introduce no static overhead, enabling uninterrupted training for larger models and increasing throughput by up to $1.46\times$ for GPT-3 Medium. Additionally, \NAME has reduced reconfiguration overhead; during failure normalization, it requires parameter migration for at most one GPU, compared to the need for reconfiguring an entire data pipeline with Oobleck. This efficiency results in up to $1.29\times$ higher training throughput. %Under frequent failures (every 30 minutes) without repairs, \NAME achieves 85\% and 66\% of the fault-free training throughput for the GPT-3 3.35B and 6.7B models, respectively. %Notably, both \NAME and Oobleck do not incur any penalty over DeepSpeed for fault-free execution, unlike Bamboo.

\paragraph{What is \NAME's throughput advantage in dynamic training scenarios?}  
We next assessed \NAME's performance in a dynamic training scenario characterized by GPU failures and re-joins, using a real-world failure trace. This trace, derived from GCP instances utilized by Bamboo~\cite{bamboo} and Oobleck~\cite{oobleck}, was replayed over a 6-hour training run on our Azure cluster. While we observed similar outcomes with Bamboo's AWS trace, we opted to omit those plots due to space limitations. During the experiment, we had 24 GPUs available instead of the originally planned 32. Figure~\ref{fig:real-world-trace}a illustrates the fluctuating number of available GPUs over time, ranging from a maximum of 24 to a minimum of 15. Unlike the previous experiment, this scenario frequently saw GPUs being removed and reintroduced into the cluster.

In Figure~\ref{fig:real-world-trace}b and~\ref{fig:real-world-trace}c, we present the training throughput achieved by Bamboo, Oobleck, and \NAME while replaying the trace for both GPT-3 Medium and GPT-3 6.7B training jobs. The solid lines indicate instantaneous training throughput, while the dashed lines represent the average training throughput for each system throughout the 6-hour period. Notably, Bamboo is unable to train GPT-3 6.7B due to memory constraints. \NAME demonstrates a performance increase of \(1.64\times\) in average throughput compared to Bamboo on GPT-3 Medium, and a \(1.46\times\) improvement over Oobleck on GPT-3 6.7B. As mentioned earlier, Bamboo incurs overhead from redundant computations, while Oobleck experiences significant stalls due to parameter re-shuffling during GPU failures and re-joins, leading to substantial drops in throughput. Additionally, during stable periods, Oobleck grapples with imbalanced heterogeneous pipelines. In contrast, \NAME consistently delivers the highest and most stable training throughput throughout the entire trace.

\subsection{\NAME Scalability}\label{sec:exp_scalability}
\begin{table}[!t]
  \captionsetup{justification=centering}
  \caption{Gap between real-world and simulated throughput across various models and failure rates.}
  \label{tbl:sim}
  \centering
  \begin{tabular}{c|cccc}
  \toprule
   %Models & \multicolumn{4}{c}{Gap} \\  \cline{1-10}
   \textbf{Models} & \textbf{Fault-Free} & \textbf{6h} & \textbf{2h} & \textbf{30m} \\ \hline
   GPT-3 Medium & -0.87\%& +5.98\%& -1.93\%& -1.48\% \\
   GPT-3 3.35B & -0.13\%& -1.58\%& +2.12\%& -1.90\% \\
   GPT-3 6.7B & +3.94\%& +2.71\%& -1.86\%& -0.85\% \\
   \bottomrule
  \end{tabular}
\end{table}

Due to the unavailability of a cluster with thousands of GPUs, we developed a simulator capable of calculating the training throughput based on a model and cluster configuration, given a specific execution plan. The simulator leverages real-world profiled statistics for each pipeline operation associated with the respective model.
To validate the simulator's accuracy, Table~\ref{tbl:sim} presents the differences between the simulated throughput and the measured real-world throughput on Azure across various failure rates for three GPT-3 models, with a maximum discrepancy of $5.98\%$. These variations are mainly due to minor fluctuations in the execution time of NCCL collectives, but they have little impact on \NAME's performance.

\begin{figure}[!t]
    \centering
    \includegraphics[width=1\linewidth]{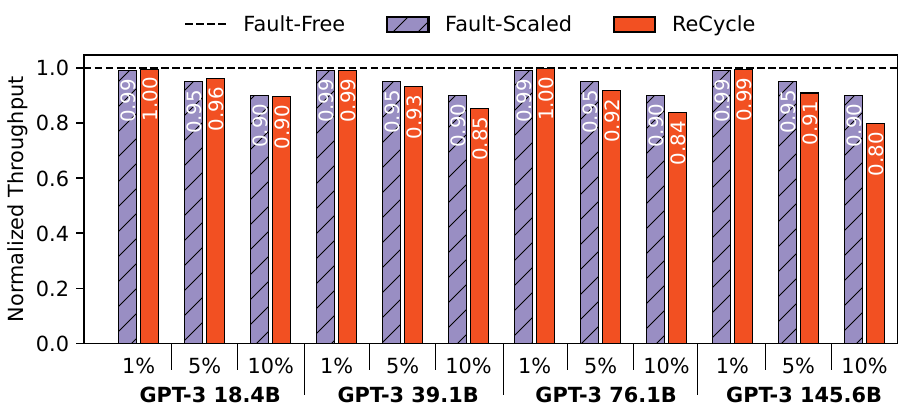}
    \caption{Simulated throughput of \NAME as model size increases, normalized to the simulated fault-free 1F1B throughput. The Fault-Scaled throughput is the fault-free throughput scaled by the percent of non-failed GPUs.}\label{fig:scalability}
\end{figure}

\paragraph{How effectively does \NAME scale to large clusters and models?}
Using the simulator, we report the training throughput for GPT models with sizes of 18.4B, 39.1B, 76.1B, and 145.6B across different clusters: (256 GPUs, 8 stages per pipeline, 32 pipelines), (512 GPUs, 16 stages per pipeline, 32 pipelines), (1024 GPUs, 32 stages per pipeline, 32 pipelines), and (1536 GPUs, 64 stages per pipeline, 24 pipelines). Figure~\ref{fig:scalability} presents the throughput normalized to the fault-free throughput achieved by DeepSpeed's 1F1B schedule. Additionally, we introduce the \textit{fault-scaled} throughput, calculated by multiplying the fault-free throughput by the percentage of operational GPUs. Notably, we focus on steady-state throughput for scenarios with 1\%, 5\%, and 10\% GPU failures, rather than the throughput during periods of GPU failures and re-joins.

\NAME demonstrates a strong capability to maintain high throughput across varying failure rates and model sizes. For a failure rate of 1\%, \NAME manages failures at or better than the fault-scaled throughput for all models and cluster sizes, frequently exhibiting no degradation in performance. The bubbles in peer workers are more than adequate to efficiently handle the workload of failed GPUs utilizing \NAME's optimizations (see Figure~\ref{fig:staggered-optimizer}). As the number of failed GPUs increases, \NAME continues to sustain high throughput levels. At a 5\% failure rate, \NAME's performance is comparable to the fault-scaled throughput. Even at a significant 10\% failure rate (e.g., 154 failed GPUs for GPT-3 145.6B), \NAME allows clusters to continue training while observing only between $0.5\%$ and $11.5\%$ degradation from the fault-scaled case. This indicates that \NAME is well-suited for training tasks in supercomputing-scale clusters with dynamic resource availability~\cite{megascale, unicorn, tpu-resiliency}.

%insupercomputing jobs for training in environments with dynamic resource availability~\cite{megascale, unicorn, tpu-resiliency}. 

\subsection{\NAME Performance Breakdown}\label{sec:eval_breakdown}
\begin{figure}
    \centering
    \includegraphics[width=1\linewidth]{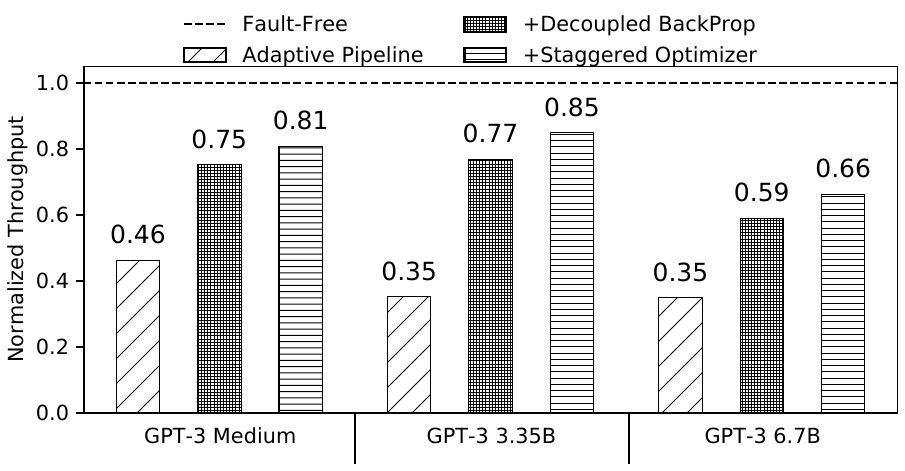}
    \caption{The contribution of the three optimization techniques to \NAME's training throughput. }
    \label{fig:impact-of-recycle-tech}
\end{figure}

\begin{figure}
    \centering
    \includegraphics[width=1\linewidth]{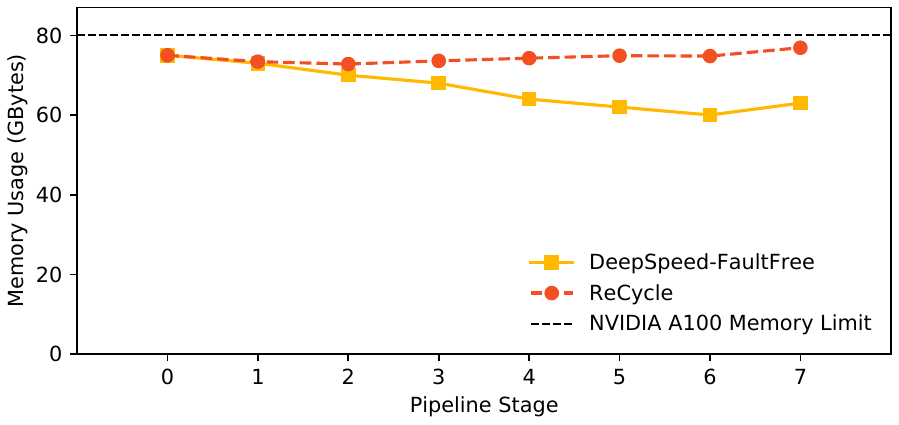}
    \caption{Memory utilization for \NAME across pipeline stages for GPT-3 6.7B with 30m failure rates compared to DeepSpeed (failure-free). The dashed line shows the 80GB memory capacity of the A100 GPU.}
    \label{fig:mem-utilization}
\end{figure}

%We next perform a series of in-depth analysis to explore the benefits and costs of each of \NAME's components.

\paragraph{How does each of \NAME's techniques contribute to performance?}

To assess the benefits and necessity of each technique in \NAME, as outlined in Section~\ref{sec:system}, we conducted an ablation study. This involved repeating the real-world experiment on the Azure cluster with GPT-3 models, using a failure frequency of 30 minutes. We adjusted the MILP formulation in the {\it Planner} to progressively enable each optimization. Figure~\ref{fig:impact-of-recycle-tech} shows the throughput achieved at each stage of the study, normalized to the fault-free throughput.

\textit{Adaptive Pipelining} allows \NAME to continue training despite failures, but it experiences significant throughput degradation due to the overhead from additional work required by peers of the failed workers. By introducing \textit{Decoupled BackProp}, \NAME boosts the training throughput by 63\% to 118\%, effectively utilizing bubbles during the cool-down phase of the pipeline schedule to mask this overhead (see Figure~\ref{fig:adaptive-schedule}). Further enhancements come from the \textit{Staggered Optimizer}, which boosts throughput by an additional $7\%$ to $11\%$ by leveraging bubbles from the warm-up phase of the pipeline schedule.

\paragraph{Can \NAME efficiently exploit unused GPU memory in hybrid parallelism?} We recognize that GPU memory utilization varies across different stages in hybrid parallelism, with later pipeline stages typically requiring less memory. To leverage this, the \textit{Decoupled BackProp} technique utilizes the available surplus GPU memory as a buffer, allowing for the postponement of $B_{weight}$ computations (see Figure~\ref{fig:split-backward-adaptive-schedule}). Figure~\ref{fig:mem-utilization} shows the peak GPU memory utilization across the eight stages of pipelined execution for \NAME compared to the fault-free DeepSpeed, based on training the GPT-3 6.7B model with a 30-minute failure interval.

DeepSpeed -- and any system utilizing a pipeline schedule akin to 1F1B -- fails to fully utilize GPU memory, especially in the later pipeline stages~\cite{pipedream}. This surplus arises because GPU workers in earlier stages must store more intermediate results, as shown in Figure~\ref{fig:fault-free-and-adaptive-schedule}. In contrast, \NAME effectively capitalizes on this opportunity, achieving near-complete utilization of GPU memory to optimize adaptive pipelined execution in the presence of failures. The \textit{Planner}'s MILP formulation is tailored to model and manage these additional memory requirements while adhering to GPU memory constraints, thereby preventing memory exhaustion.

\begin{figure}[!t]
    \centering
    \includegraphics[width=1\linewidth]{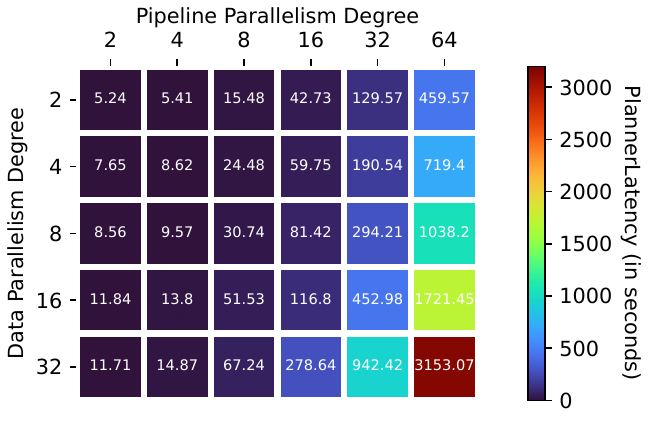}
    \caption{\NAME \textit{Planner} latency (in seconds) to find optimized schedules for up to $25\%$ of failed GPUs.}\label{tbl:planner}
\end{figure}

\begin{comment}
\begin{table}[!t]
  \captionsetup{justification=centering}
  \caption{\NAME \textit{Planner} latency (in seconds) to find optimized schedules for up to $25\%$ of failed GPUs.}
  \label{tbl:planner}
  \centering
  \begin{tabular}{c|cccccc}
  \toprule
   DP & \multicolumn{6}{c}{PP Degree} \\
   Degree & 2 & 4 & 8 & 16 & 32 & 64 \\ \hline
    2 &	5.24&	5.41&	15.48&	42.73&	129.57&	459.57 \\
    4 &	7.65&	8.62&	24.48&	59.75&	190.54&	719.4 \\
    8 &	8.56&	9.57&	30.74&	81.42&	294.21&	1038.2 \\
    16 & 11.84&	13.8&	51.53&	116.8&	452.98&	1721.45 \\
    32 & 11.71&	14.87&	67.24&	278.64&	942.42&	3153.07 \\
   \bottomrule
  \end{tabular}
\end{table}
\end{comment}

\paragraph{What is the overhead of \NAME's \textit{Planner}?}  
To evaluate the latency of the \textit{Planner} in generating adaptive schedules for large clusters, Figure~\ref{tbl:planner} shows the time required to generate all necessary schedules for up to 25\% node failures across various hybrid parallelism strategies. Users have the flexibility to configure the fault-tolerance threshold to suit their needs, whether smaller or larger. The reported latency encompasses both phases of the \textit{Planner} and was measured using the Gurobi MILP solver~\cite{gurobi} on a 96-core CPU.

In a training setup with 2048 GPUs (with $DP=32$ and $PP=64$), the \textit{Planner} can generate all adaptive schedules to accommodate up to $512$ failures in just $3153$ seconds (or $52.5$ minutes). Considering that training large models on thousands of GPUs typically takes weeks~\cite{megatron-nlg, megascale, training-at-meta}, the \textit{Planner}'s latency is negligible, accounting for less than $0.1\%$ of the overall training time.

\multilinecomment{
We seek to answer following questions:

\subsection{Simulator vs Real-World} X-Axis: Model Size, Y-Axis: Samples/sec. Legend: Azure, Simulator.
\subsection{Fault Free Performance}
\myzhao{Rename: Is slipstream efficient with handling num errors.. (or something like that)}
\subsection{How does \NAME scale with number of failures?} X-Axis: Number of Failures, Y-Axis: Samples/sec. Model size and cluster size is fixed.

\subsection{How does \NAME scale with cluster size?} X-Axis: Model Size (130B to 1T), Y-Axis: Samples/sec. Number of failures is fixed.

\subsection{Does \NAME affect model quality?} X-Axis: Iteration, Y-Axis: Loss

\subsection{How does \NAME compare to other systems (Oobleck, Bamboo, Varuna)?} X-Axis: Fault tolerance techniques/systems, Y-Axis: Samples/sec. 3-plots: One per model size, number of faults is fixed.

\subsection{Ablation}
    \paragraph{Memory overhead of Split Backward} A table, reporting memory usage of unified backward and split backward.
    \paragraph{Compute overhead of \NAME planner?} A table, reporting planner overhead as a function of cluster size.
    \paragraph{Overhead of \NAME in fault-free case} X-Axis: Model Size, Y-Axis: Samples/sec

\subsection{Does \NAME scale with model size/type and schedules?}
    \paragraph{Performance for shallow and wide models} 
    \paragraph{Performance for deep and narrow models}
    \paragraph{\NAME support for other schedules? (e.g.; Chimera, kFkB,...)} X-Axis: Number of failures, Y-Axis: Samples/sec. Legend: Pipeline schedules
}
\section{Related Work}\label{sec:related}

{\bf Parallel Training.} Data parallelism~\cite{imagenet, sgd, dean-2012} is a widely utilized mode of parallelism that distributes the dataset across different partitions for processing. In this mode, the learned weights are synchronized via either an all-reduce approach~\cite{sgd} or by using parameter servers~\cite{parameter-server, project-adam, byteps}. Alternatively, model parallelism~\cite{one-weird-trick, megatron-lm, mesh-tensorflow} involves distributing the components of a deep neural network (DNN) model across multiple GPU devices, allowing each device to handle a specific subset of the model’s parameters for all input data. Recently, pipeline parallelism~\cite{gpipe, pipedream, chimera, hanayo, bpipe} has emerged as a technique for training large models by dividing the model's layers among different workers and utilizing micro-batches to efficiently use the available computational resources. Prominent deep learning training frameworks like PyTorch~\cite{pytorch}, DeepSpeed~\cite{deepspeed}, and Megatron~\cite{megatron-sc} have adopted hybrid-parallelism, an approach that integrates data parallelism, model parallelism, and pipeline parallelism. This integration facilitates training on a massive scale while enhancing computational and memory efficiency. Additionally, DeepSpeed introduces ZeRO-style~\cite{zero, zero-infinity, zero-offload, pytorch-fsdp, activation-checkpointing} data parallelism, which strategically partitions model states across GPUs, coordinating via communication collectives to synchronize parameters as needed.

\noindent\textbf{Optimizing Hybrid-Parallel Training.} Improvements in hybrid parallelism are extensively studied in the context of ML training~\cite{megatron-sc, activation-checkpointing, deepspeed, pytorch-distributed, pytorch-fsdp, alpa, gpipe, pipedream, chimera, hanayo, bpipe}.
Alpa~\cite{alpa} automatically optimizes inter- and intra-operator parallelism using a hierarchical ILP formulation.
Tofu~\cite{tofu} employs dynamic programming to optimally partition tensor operations in a single node.
FlexFlow~\cite{flexflow} uses a randomized search algorithm to quickly find parallelism strategies.
TensorOpt~\cite{tensoropt} introduces a dynamic programming approach capable of optimizing parallelism across multiple resource dimensions, including memory and compute.
Piper~\cite{piper} proposes a two-level dynamic programming algorithm to find optimal hybrid parallelism strategies. 
In contrast, \NAME leverages a dynamic programming algorithm and mixed-integer linear programming (MILP) to exploit key functional redundancies in hybrid parallelism, explicitly enhancing training throughput in the presence of failures.

\noindent{\bf Elastic Training.} Elastic training systems can dynamically adjust the resources allocated to a training job. Both Horovod \cite{horovod} and Torch Distributed~\cite{pytorch-distributed} offer mechanisms to modify the number of workers, but require either a restart from a checkpoint or an expensive re-shuffling of model parameters. CoDDL~\cite{coddl}, Optimus~\cite{optimus}, OASiS~\cite{oasis}, and Themis~\cite{themis} are ML cluster schedulers that can dynamically allocate resources across multiple DNN training jobs. Or et al.~\cite{mlsys20:or} auto-scale the number of workers for a training job. 
Varuna~\cite{varuna} provides elastic training by leveraging spot instances, but requires restarts from checkpoints to handle unexpected preemptions. While some elastic training systems can affect model consistency by changing critical hyperparameters such as batch size and learning rate, \NAME guarantees mathematical consistency of operations, regardless of the number of failures.

\noindent{\bf Fault-tolerant Training.} DNN training systems commonly use checkpoints for fault recovery~\cite{deepspeed, tpu-resiliency, bloom, megascale, unicorn, laion, microsoft-training-trace, lineage}, though naive checkpointing can cause stalls. Recent works like CheckFreq~\cite{checkfreq}, Check-N-Run~\cite{check-n-run}, and Gemini~\cite{gemini} reduce overheads by adapting checkpoint frequency, quantizing embedding tables, and scheduling checkpoint traffic across the storage hierarchy, respectively. Megascale~\cite{megascale} alleviates the storage bottleneck during recovery by sharing data between corresponding GPU workers across data parallel groups. 

Two recent projects have proposed to utilize pipeline parallelism for efficient training in the presence of failures without spares. Bamboo~\cite{bamboo} introduces redundant computation (RC) to provide resilience in the presence of frequent preemptions of training with spot instances. Oobleck~\cite{oobleck} provides resiliency through creation of heterogeneous pipelines. \NAME matches or outperforms them for a wide range of model sizes and failure frequencies.
\section{Conclusion}\label{sec:conclusion}
\NAME enables efficient DNN training in the presence of failures without relying on spare resources. By leveraging the functional redundancies inherent in hybrid parallel training systems for large DNNs, \NAME introduces innovative techniques to minimize the degradation in training throughput caused by failures. It does this by utilizing unused resources, such as pipeline bubbles. We evaluated an end-to-end prototype of \NAME and demonstrated its ability to tolerate a high number of concurrent failures across systems and models of varying sizes. Compared to Oobleck and Bamboo, \NAME improves training throughput under failure conditions by up to $1.46\times$ and $1.64\times$ respectively.

\section*{Acknowledgements}
We are grateful to the anonymous reviewers and to our shepherd, Laurent Bindschaedler, whose comments have greatly helped improve this paper. This research was partly supported by the Stanford Platform Lab and its affiliates, and by ACE, one of the seven centers in JUMP 2.0, a Semiconductor Research Corporation (SRC) program sponsored by DARPA. Athinagoras Skiadopoulos was supported by a Stanford Graduate Fellowship. Mark Zhao was supported by a Stanford Graduate Fellowship and a Meta PhD Fellowship.

\bibliographystyle{ACM-Reference-Format}
\bibliography{references}

% New page after bibs. We need to move appendix to separate pdf
% \clearpage
% Appendix
% \input{sections/appendix}

\end{document}